\documentclass[12pt,reprint,showpacs,showkeys,preprintnumbers,nofootinbib,aps,amsmath,amssymb]{revtex4-1}

\usepackage{xcolor}
\usepackage{amsmath}
\usepackage{amsthm}
\usepackage{amssymb}
\usepackage{graphicx}
\usepackage{hyperref}
\usepackage{url}
\usepackage{cleveref}
\usepackage[caption = false]{subfig}
\def\<#1>{\langle\ignorespaces#1\unskip\rangle}
\usepackage{dcolumn}
\usepackage{bm}

\begin{document}
\title{Three-state opinion dynamics in modular networks}

\author{Andr\'{e} L. Oestereich $^{1}$}
\thanks{andrelo@id.uff.br}

\author{Marcelo A. Pires $^{2}$}
\thanks{piresma@cbpf.br}

\author{Nuno Crokidakis $^{1}$}
\thanks{nuno@mail.if.uff.br}

\affiliation{$^{1}$Instituto de F\'{\i}sica, Universidade Federal Fluminense, Niter\'oi/RJ, Brazil \\ 
$^{2}$Centro Brasileiro de Pesquisas F\'isicas, Rio de Janeiro/RJ, Brazil}

\date{\today}

\begin{abstract}
In this work we study the opinion evolution in a community-based population with intergroup interactions.  We address two issues.  First, we consider that such intergroup interactions can be negative with some probability $p$.  We develop a coupled mean-field approximation that still preserves the community structure and it is able to capture the richness of the results arising from our Monte Carlo simulations: continuous and discontinuous order-disorder transitions as well as nonmonotonic ordering for an intermediate community strength.  In the second part, we consider only positive interactions, but with the presence of inflexible agents holding a minority opinion. We also consider an indecision noise: a probability $q$ that allows the spontaneous change of opinions to the neutral state. Our results show that the modular structure leads to a nonmonotonic global ordering as $q$ increases. This inclination toward neutrality plays a dual role: a moderated propensity to neutrality helps the initial minority to become a majority, but this noise-driven opinion switching  becomes less pronounced if the agents are too susceptible to become neutral.
\end{abstract}


\maketitle

\section{Introduction}
\label{intro}

What are the requirements for the upraise of consensus or polarization is one of the main questions of sociophysics \cite{2009castellanoFL,2012galam,2008galam}. This field consists of the application of statistical physics methods to the study of social systems. In order to answer this question several models of opinion were already proposed.

Although the use of continuous models \cite{2017panQXTH,2014javarone,2015javaroneS,2012crokidakisA,2014terranovaRS,2011biswas,2016calvaoRA,2016mukherjeeC,2015ramosSRA,2016vieiraC} enables the modelling of broader social contexts, there are many social scenarios  in which the possible choices are limited and thus can be modeled by discrete variables \cite{2000deffuantNAW,2007lorenz,2008martins,2010lallouacheCCC,2011biswasCC,2016vieiraAC, 2017anteneodoC} as was done in this work. Apart from this, discrete models have the advantage of allowing a better understanding of the underlying mechanism behind the macroscopic outcomes through an analytical treatment.

A simple rule for the evolution of both discrete and continuous models, that has been considered previously~\cite{2012biswasCS}, is

\begin{equation}
    o_i (t+1) = o_i(t) +\mu_{i j} o_j(t),
    \label{kinnEx}
\end{equation}

\noindent with $1 \leq i, j \leq N$, where $N$ is the population size, $i \neq j$, and $\mu_{ij}$ are the coupling coefficients. These coefficients dictate if the opinion of the $j$-th agent influences the $i$-th agent's opinion at the time $t+1$. Hence, the coefficients $\mu_{ij}$ can be viewed as an adjacency matrix, where $\mu_{ij} = 0$ if the individuals $i$ and $j$ are not connected, and $\mu_{ij} = 1$ if they are connected.

Since the right-hand side of Eq. (\ref{kinnEx}) can exceed the extreme values ($\pm 1$), it is also necessary to forbid changes in opinions that exceed the limiting values. Or, equivalently, to reinsert the opinion back to its corresponding limiting value. This additional rule introduces nonlinearity into the system's evolution.

This model has been extensively studied in several different networks, but not yet in networks that exbibit modular structures. Modular structures have been found in many real-world social and biological networks \cite{2006newman,2010fortunato}. These networks present much more dense links within modules than those among modules. Many previous studies have shown that this structure has a significant impact on the dynamics taking place on networks such as synchronization, \cite{2006arenasDPC,2008liLAS} epidemic \cite{2005liuH} or information spreading \cite{2006huangPL,2014nematzadehFFA}, opinion formation \cite{2007lambiotteA,2007lambiotteAH,2008ruL,2009siLZ,2015fengHC} and Ising-like phase transitions \cite{2009panS,2009dasguptaPS,2009sucheckiH,2011chenH}.

In this work we consider the kinetic opinion dynamics in modular networks. Such an approach seems even more important given the context of political discussions in social media. It was shown that in the discussion on Twitter leading the 2010 USA congressional midterm election the retweet network formed two distinct communities \cite{2011conoverRFG}. A similar community structure was observed in a political communication network constructed based on users that interchanged opinions related to the impeachment of former Brazilian President Dilma Rousseff \cite{2019cotaFPS}. It was also shown that in an abortion discussion replies between different-minded individuals reinforce in-group and out-group affiliation \cite{2010yardiB}.

More specifically we address two issues. In the first problem, the main difference among our model and the models presented so far is that we consider intergroup bias. This is relevant because people have shown in-group favoritism and out-group derogation \cite{2002hewstoneRW}. This behavior has been shown to arise when individuals differ in some critical but unobservable way and this difference is associated to some symbolic marker \cite{2008effersonLF}.  In the second part, we treat the question of how the multifold interplay between modular structure,  noise towards neutrality and peer-pressure impacts on the minority spreading of a localized opinion of inflexible agents. This is an important issue for social dynamics \cite{2002galam,2017biswasS,2007galamJ}.


\section{Framework}
\label{model}

\subsection{Generating the network}
\label{genNet}

First of all, it is important to define the community structure because the interactions dependent on it. To systematically investigate the impact of community structure, we prepare an ensemble of networks with two communities with a varying degree of strength, using the block-model approach \cite{2002girvanN,2008lancichinettiFR,2011karrerN,2014nematzadehFFA}. Another approach for building a network with community structure can be found in \cite{2018javaroneM}.

\begin{figure}[h!]
    \centering
    \subfloat[]{\includegraphics[width=0.20\textwidth]{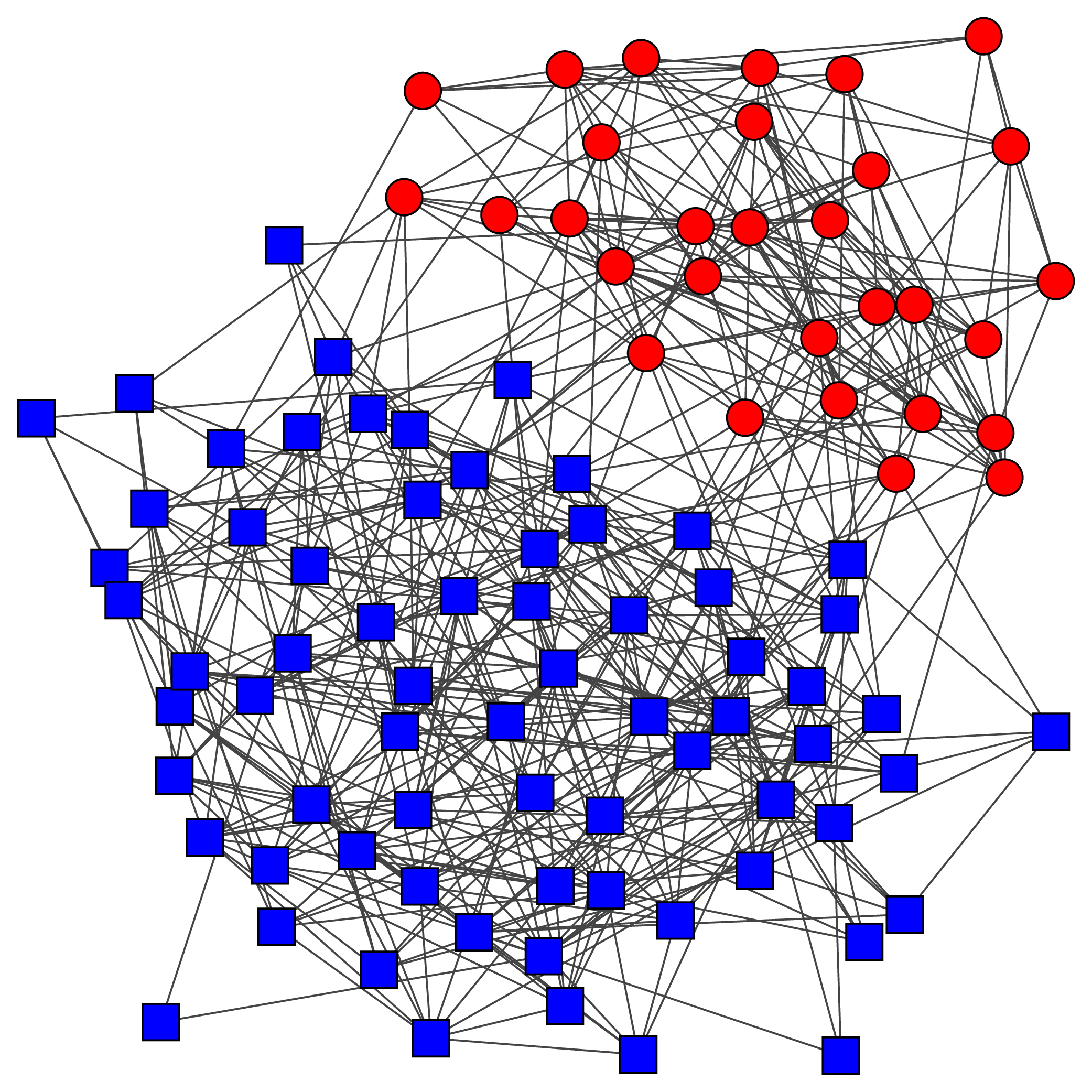}} \quad
    \subfloat[]{\includegraphics[width=0.20\textwidth]{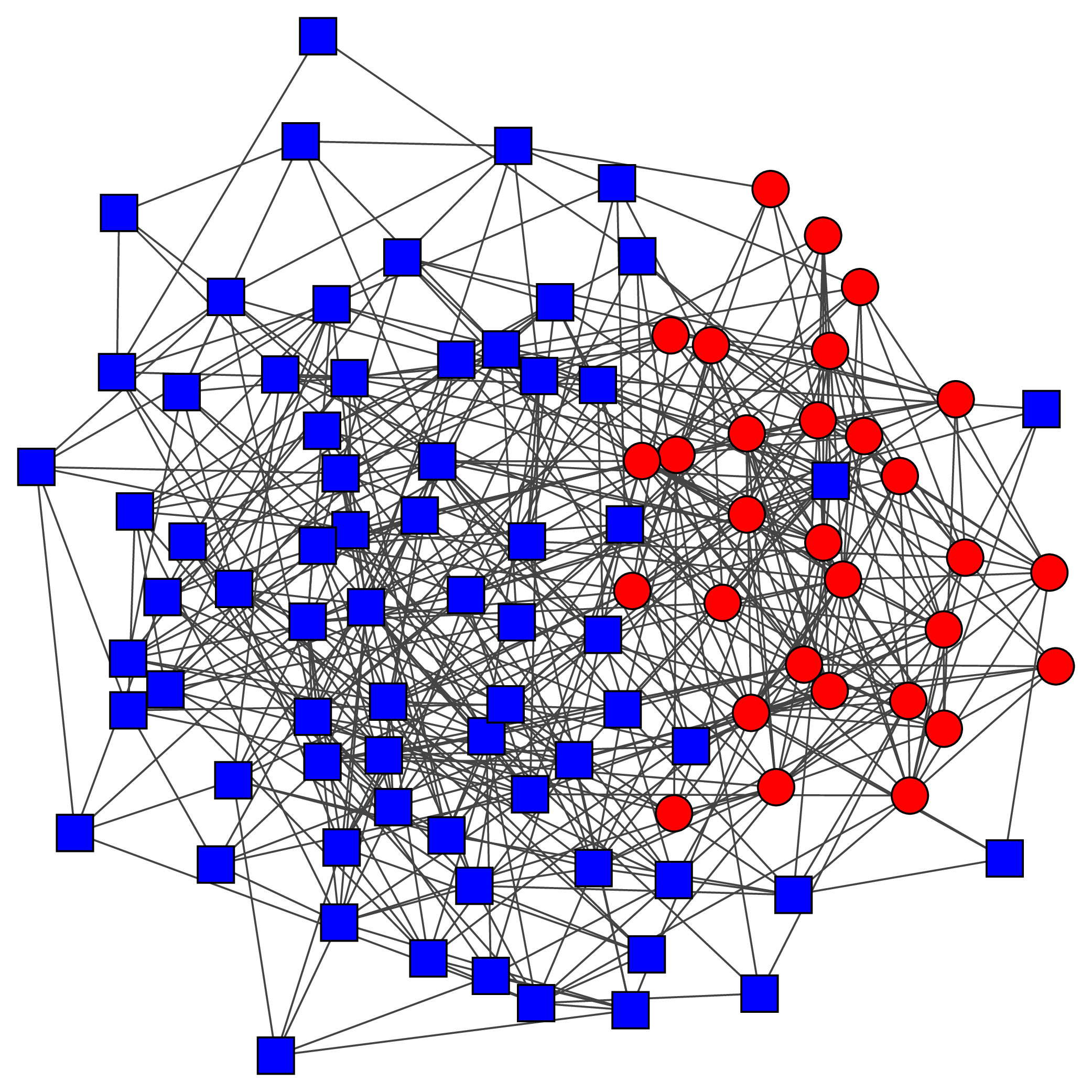}} \\
    \subfloat[]{\includegraphics[width=0.20\textwidth]{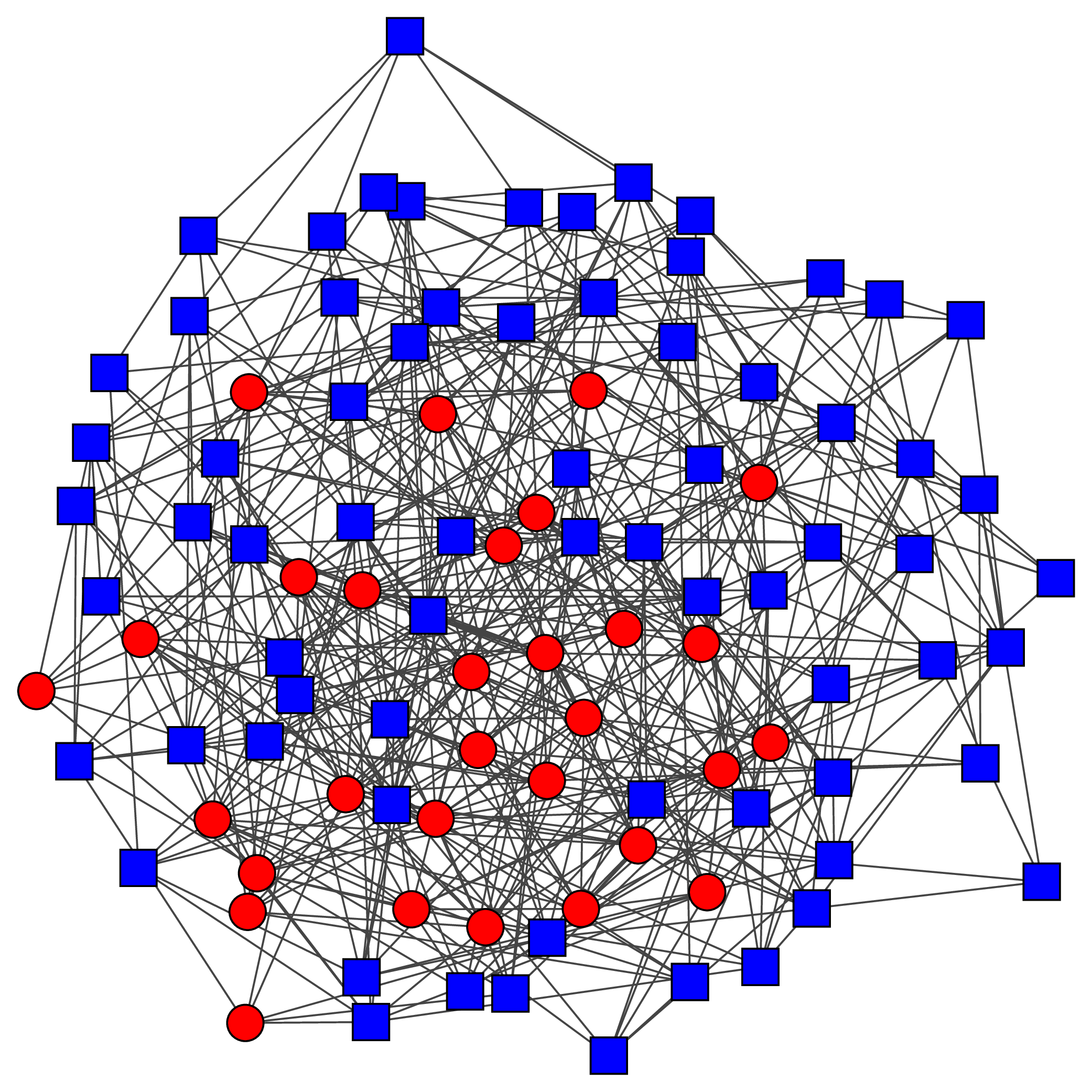}} \quad
    \subfloat[]{\includegraphics[width=0.20\textwidth]{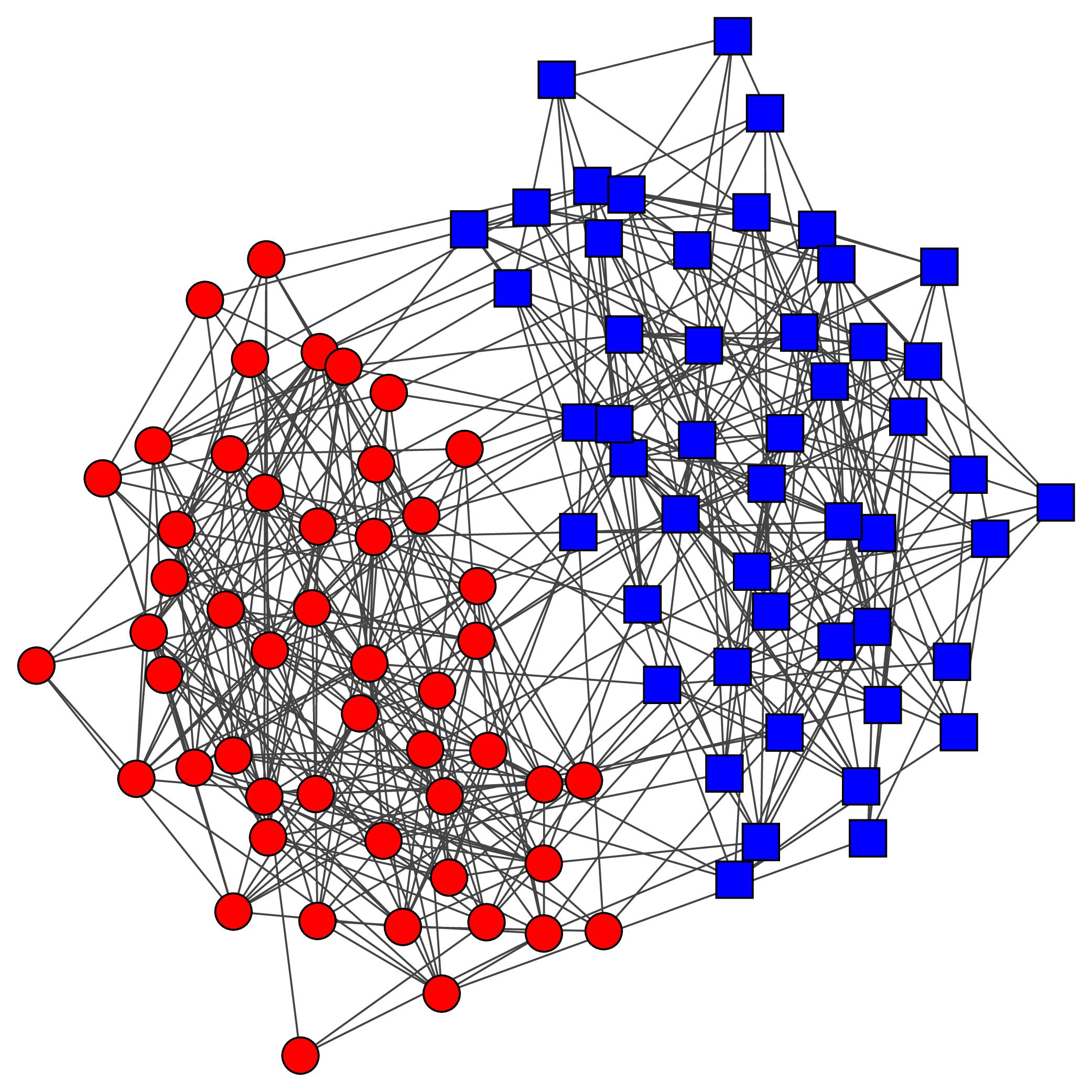}} \\
    \subfloat[]{\includegraphics[width=0.20\textwidth]{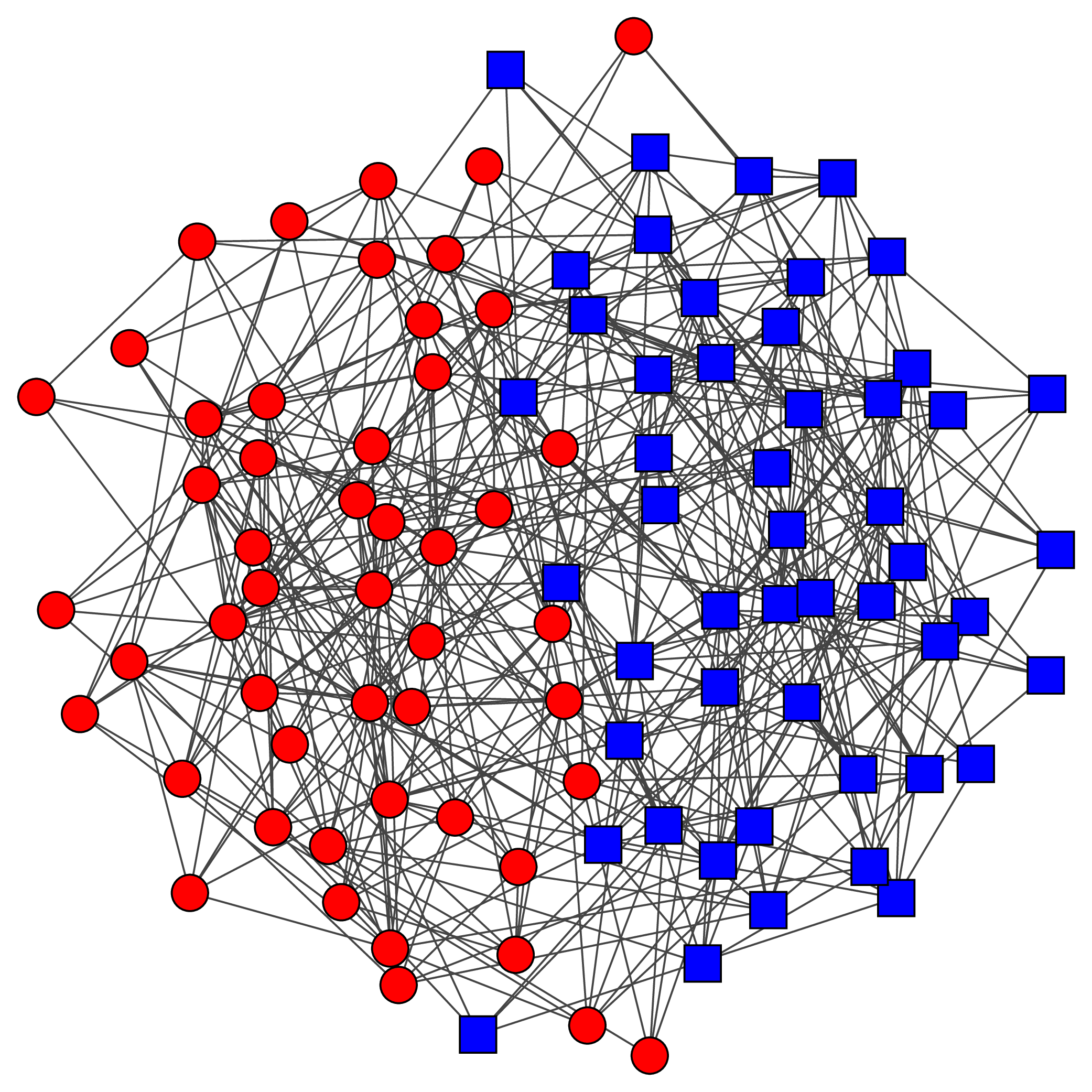}} \quad
    \subfloat[]{\includegraphics[width=0.20\textwidth]{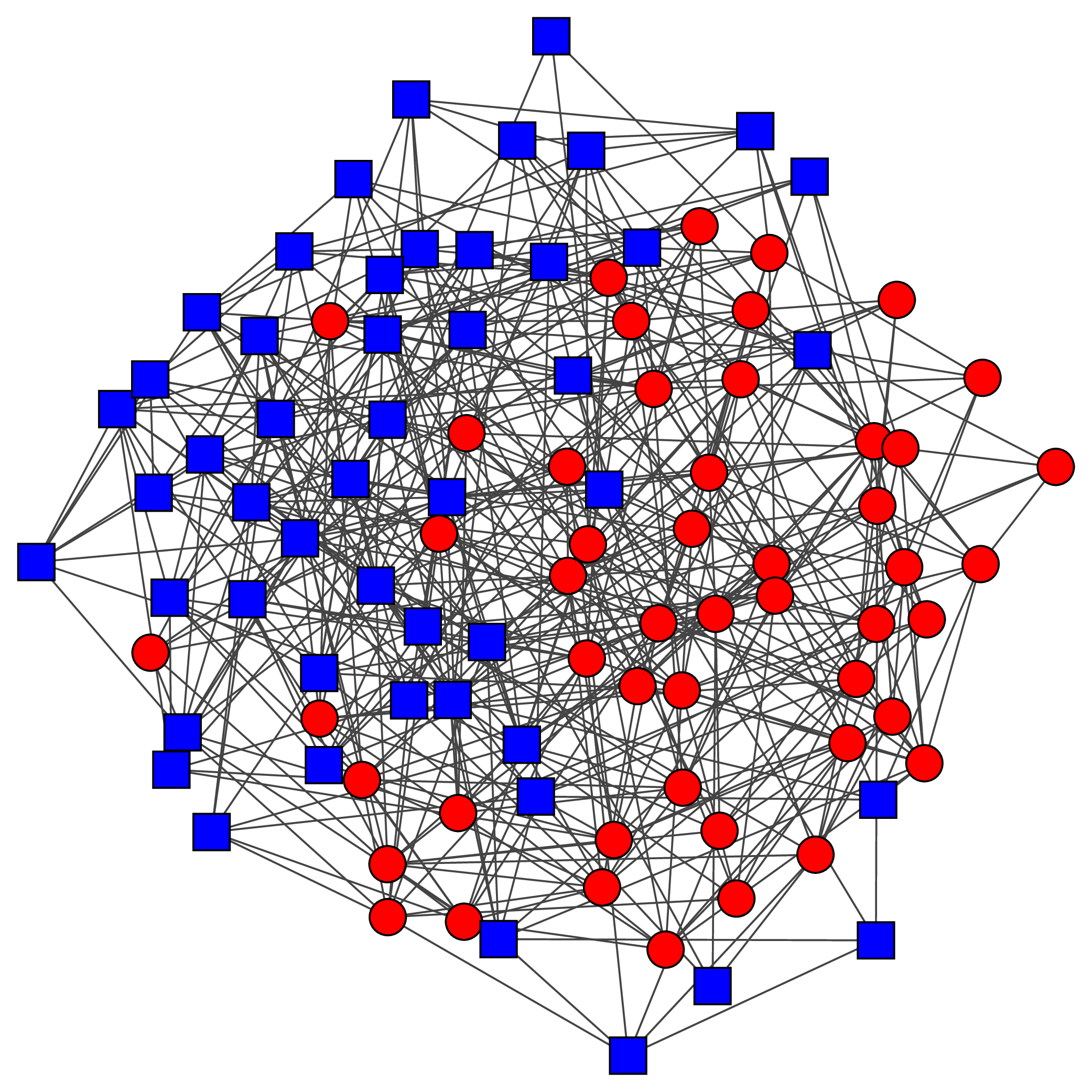}}
    \caption{Examples of the layout of the modular networks, where the blue squares represent community $1$ and the red circles the community $2$, for $N=100, \<k> = 10$ and $n_1=0.3, h=0.1$ in (a), $n_1=0.3,h=0.2$ in (b), $n_1=0.3,h=0.3$ in (c), $n_1=0.5,h=0.1$ in (d), $n_1=0.5,h=0.2$ in (e) and $n_1=0.5,h=0.3$ in (f).}
    \label{netws}
\end{figure}

We start by randomly selecting $N_1$ of the $N$ nodes and assigning them to community $1$, and assigning the other $N_2=N - N_1$ nodes to community $2$. Then, $(1-h)M$ links are randomly distributed among pairs of nodes in the same community and $h M$ are randomly distributed among pairs of nodes that belong to different communities, where $M = N\<k>/2$ is the total number of links in the whole network (see \cref{netws}). The parameter $h$ controls the strength of the community structure: a large value of $h$ yields more links between the two communities and, thus a weaker community structure.

\subsection{Fractions of in and out group connections}

Since in the kinetic exchange opinion model the agents interact with one of their neighbors at random, it will be useful to find the fractions of in- and out-group connections to perform our approximations latter. These fractions will determine the probabilities of in and out group interactions.

The whole network has $M = N \<k> / 2$ links and $z_i$  links within community $i$, such that $\<k>N_i= 2z_i + hM$. Therefore, the fraction of connections of an agent in community $i$ with a node of the same community is given by

\begin{equation}
    P (i,i) = \frac{2z_i}{N_i\<k>} = \frac{N_i\<k> - hN\<k>/2}{N_i \<k>} = 1 - \frac{hN}{2N_i} = 1 - \frac{h}{2n_i},
    \label{zii}
\end{equation}

\noindent where $n_i = N_i /N$ is the fraction of nodes in community $i$. The fraction of connections with nodes of the other community is given by

\begin{equation}
    P (i,j) = \frac{hM}{N_i\<k>} = \frac{hN\<k>/2}{N_i \<k>} = \frac{hN}{2n_i} = \frac{h}{2n_i}.
    \label{zij}
\end{equation}


\subsection{Different connectivities in each community}

One may think that assuming that both communities have the same average connectivity is an unreasonable assumption. Our results can be extended considering an effective community size. Here we will see how in this formulation of the network a community with higher connectivity is mathematically equivalent to a larger community where both communities have the same connectivity.

Let each community have $N_i k_i$ connections, where $k_i$ is the average connectivity on community $i$. We have $m$ of those connections are between communities, so in each community we have $m = Mh = h(k_1N_1+k_2N_2)/2$ connections to the other community.  Therefore, the probability of an agent interacting with another agent in the same community is given by

\begin{equation}
\begin{split}
    P(i,i) = \frac{k_i N_i -m}{k_iN_i} = 1 - \frac{h}{2} - \frac{hk_jN_j}{2k_iN_i} \\
    = 1 - \frac{h}{2} \left(1+ \frac{k_jN_j}{k_iN_i} \right) = 1 - \frac{h}{2n_i'},
\end{split}
\end{equation}

\noindent and the probability of interacting with an agent from the other community is given by

\begin{equation}
    P(i,j) = \frac{m}{k_iN_i} = \frac{h}{2} + \frac{hk_jN_j}{2k_iN_i} = \frac{h}{2} \left(1+ \frac{k_jN_j}{k_iN_i} \right) = \frac{h}{2n_i'}.
\end{equation}

\noindent Now we have an ``effective relative size'' $n_i'$ defined as

\begin{equation}
    n_i' = \frac{k_iN_i}{k_1N_1 + k_2N_2} = \frac{k_iN_i}{2M}.
\end{equation}

From this we can see that having one community more connected than another just changes its ``effective relative size'' and does not change the form of results previously presented.


\subsection{The interactions}

For both the models we consider a discrete opinion model in which each agent $i$ can have opinion $o_i =  +1, 0$ or -1. Opinions $o_i=\pm1$ are decided agents and $o_i=0$ represents an undecided or neutral agent. We have considered populations of size $N=10^4$ distributed in a network described in the previous section. As a measure of time we define a Monte Carlo step (mcs) as an update of the opinion of each one of the $N$ agents.

To characterize the coherence of the collective state of each community we consider

\begin{equation}\label{eq_m_i}
    m_i = \frac{1}{N_i} \left| \sum_{j \in \mathcal{C}_i} o_j \, \right|~, 
\end{equation}

\noindent where $\mathcal{C}_i$ is the set of individuals in community $i$ and $N_i$ the number of agents in community $i$. In this way the global order parameter is given by

\begin{equation}
    O = \frac{N_1m_1 + N_2 m_2}{N} = \frac{1}{N} \left| \sum_{j=1} ^{N} o_j \, \right|~,
    \label{oParam}
\end{equation}

\noindent where the sum is taken over both communities. Note that the time dependency is implicit.


\section{Model A: intergroup bias}

In the first formulation of our model we consider the presence of both negative and positive pairwise interactions. The negative interactions only occur between members of distinct communities, thus introducing a bias in the dynamics.

\subsection{Description}

At time step $t$ each agent (that will be referred to as $i$) updates its opinion interacting with one of its neighbors (that will be referred to as $j$), chosen at random in each time step, in one of two ways. If both agents belong to the same community they always interact positively according to

\begin{equation}
    o_i(t+1) = o_i(t) + o_j(t)
    \label{int_pos}
\end{equation}

\noindent In this case $\mu_{i j}$ of \cref{kinnEx} is simply the adjacency matrix of the network which does not change during the simulation. If agents $i$ and $j$ belong to different communities they can interact negatively with a probability $p$ according to

\begin{equation}
    o_i(t+1) = o_i(t) - o_j(t),
    \label{int_neg}
\end{equation}

\noindent and with complementary probability ($1-p$) they interact positively as in \cref{int_pos}. This differentiation in the way the agents interact with agents of the opposite communities introduces the in-group bias in our model.


\subsection{Results and discussion}

\begin{figure*}[t]
    \centering
    \subfloat[]{\includegraphics[width=0.40\textwidth]{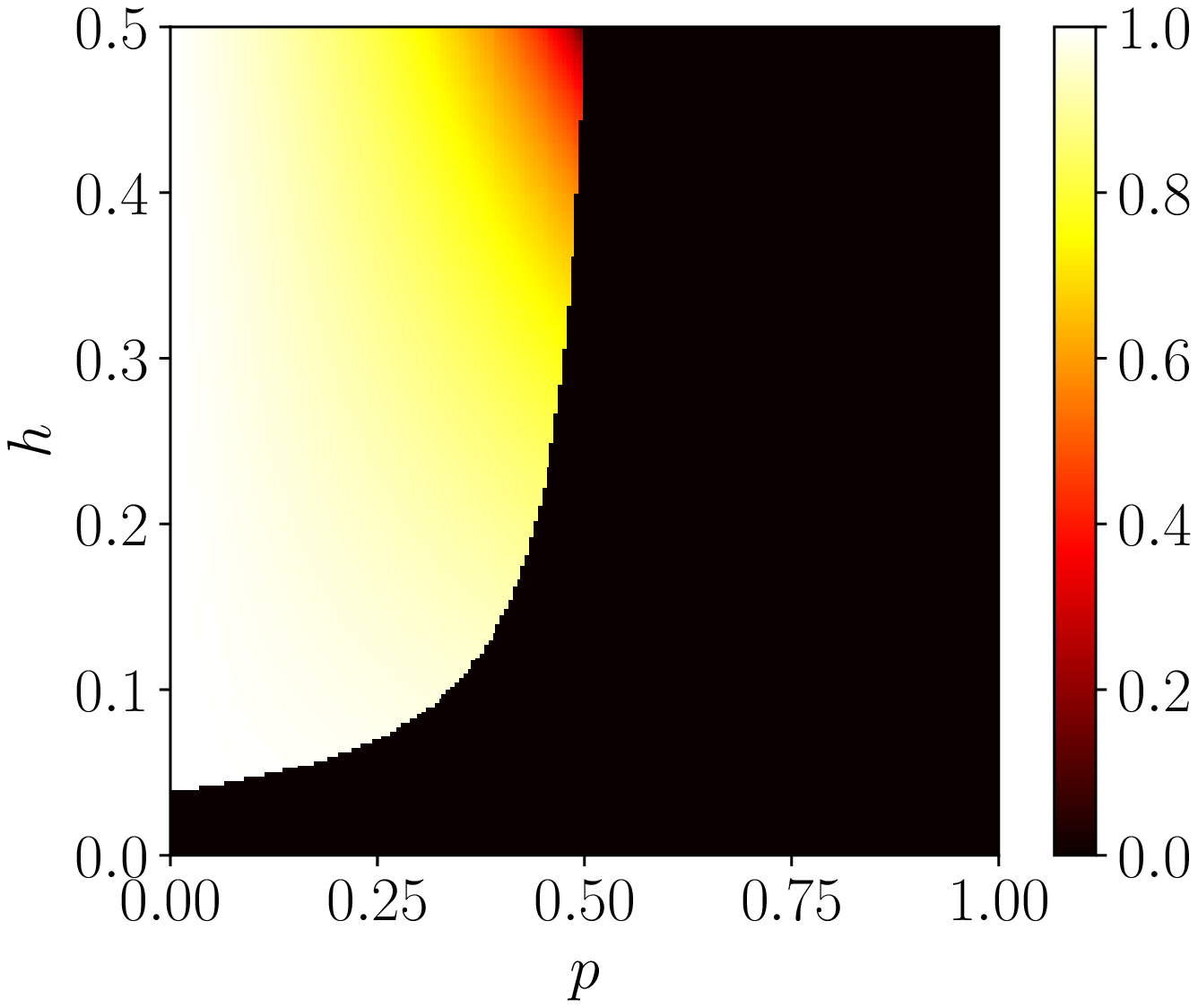}} \qquad
    \subfloat[]{\includegraphics[width=0.40\textwidth]{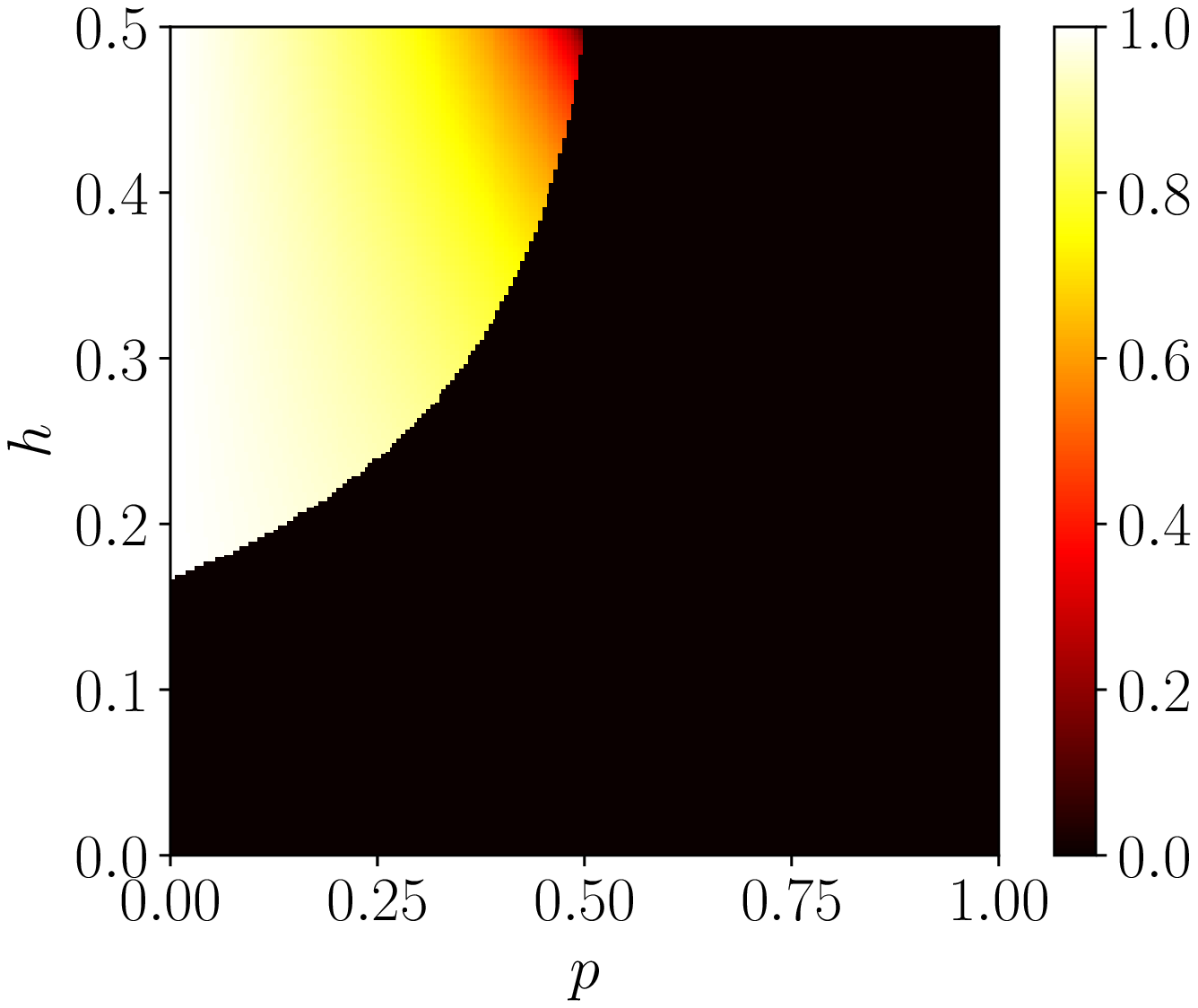}}\label{EU_Oxpxh_F11} \\
    \subfloat[]{\includegraphics[width=0.40\textwidth]{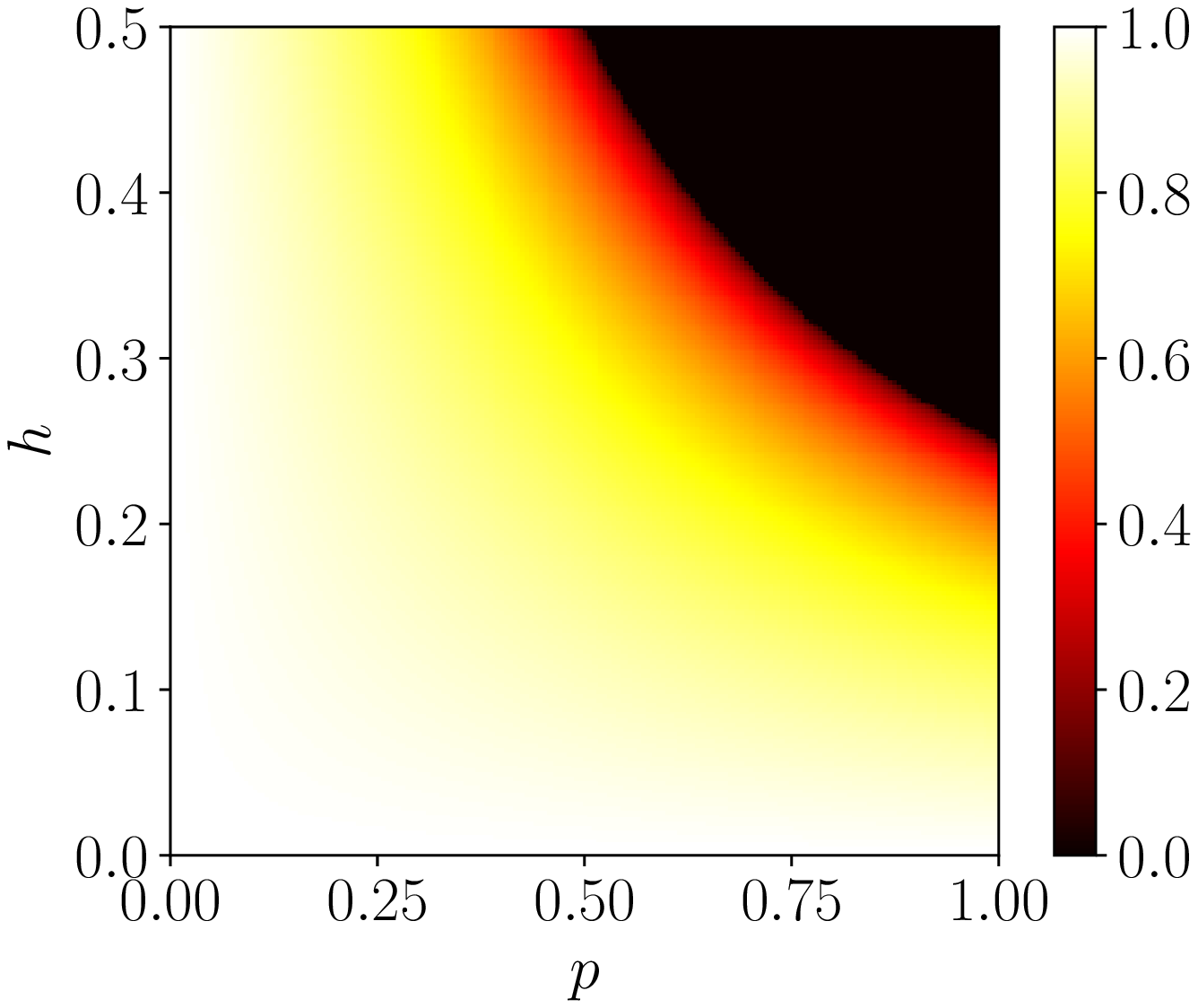}} \qquad
    \subfloat[]{\includegraphics[width=0.40\textwidth]{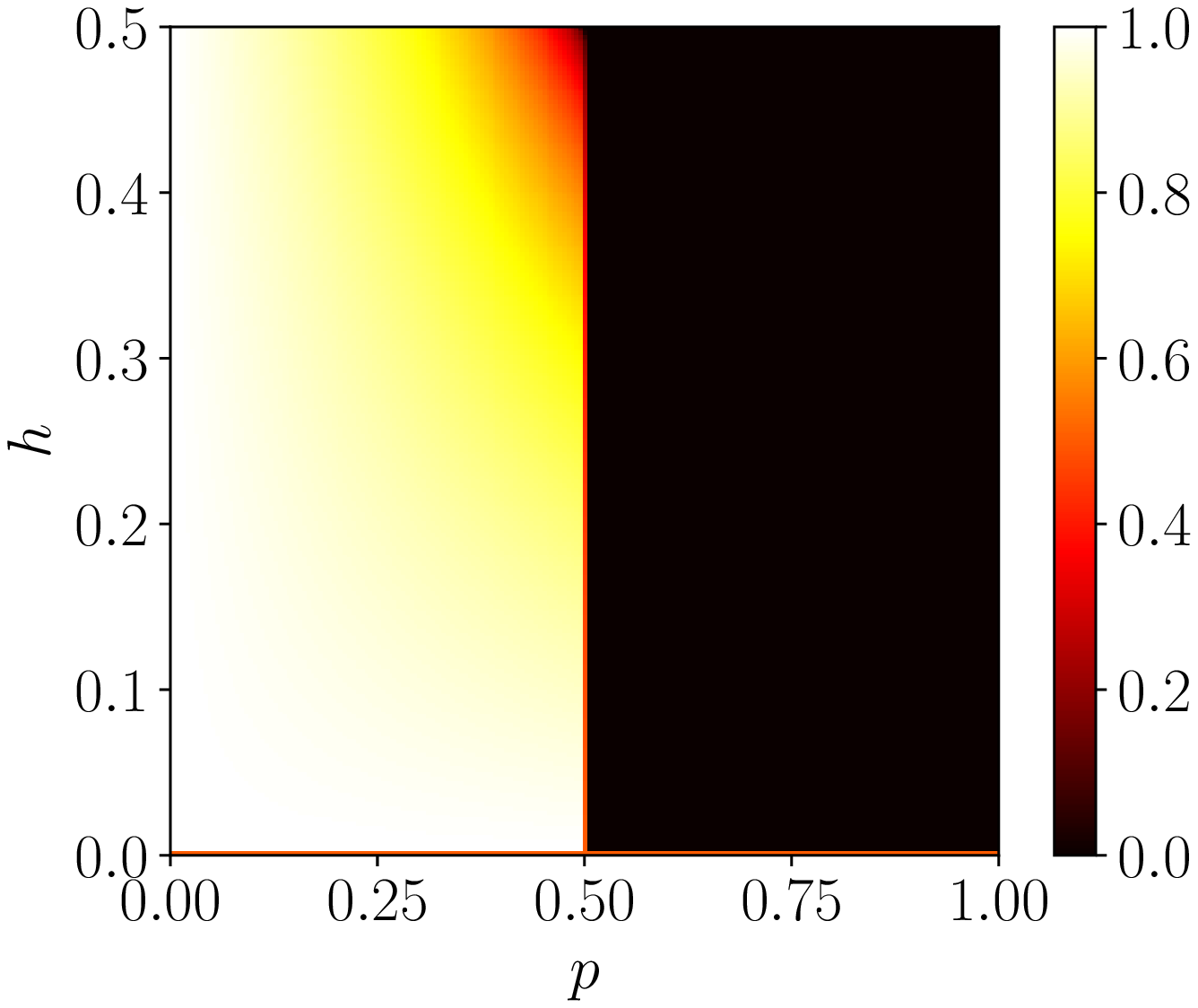}}
    \caption{Global order $O$ varying both parameters $p$ and $h$ for the numerical integration of the \cref{da,du,db} with $n_1 = 1/2$. The difference among the graphics is the initial condition. $m_1=1$ and $m_2=-1$ for (a); $m_1=0.02$, $u_1=0$, $m_2=0.0$ and $u_2 = 0.01$ for (b); $m_1=m_2=1$ for (c); $m_1=0.04$, $u_1=0.33$, $m_2=0.02$ and $u_2 = 0.33$ for (d).}
    \label{EU_OxHxP}
\end{figure*}

\begin{figure}[h]
    \centering
    \subfloat[]{\includegraphics[width=0.45\textwidth]{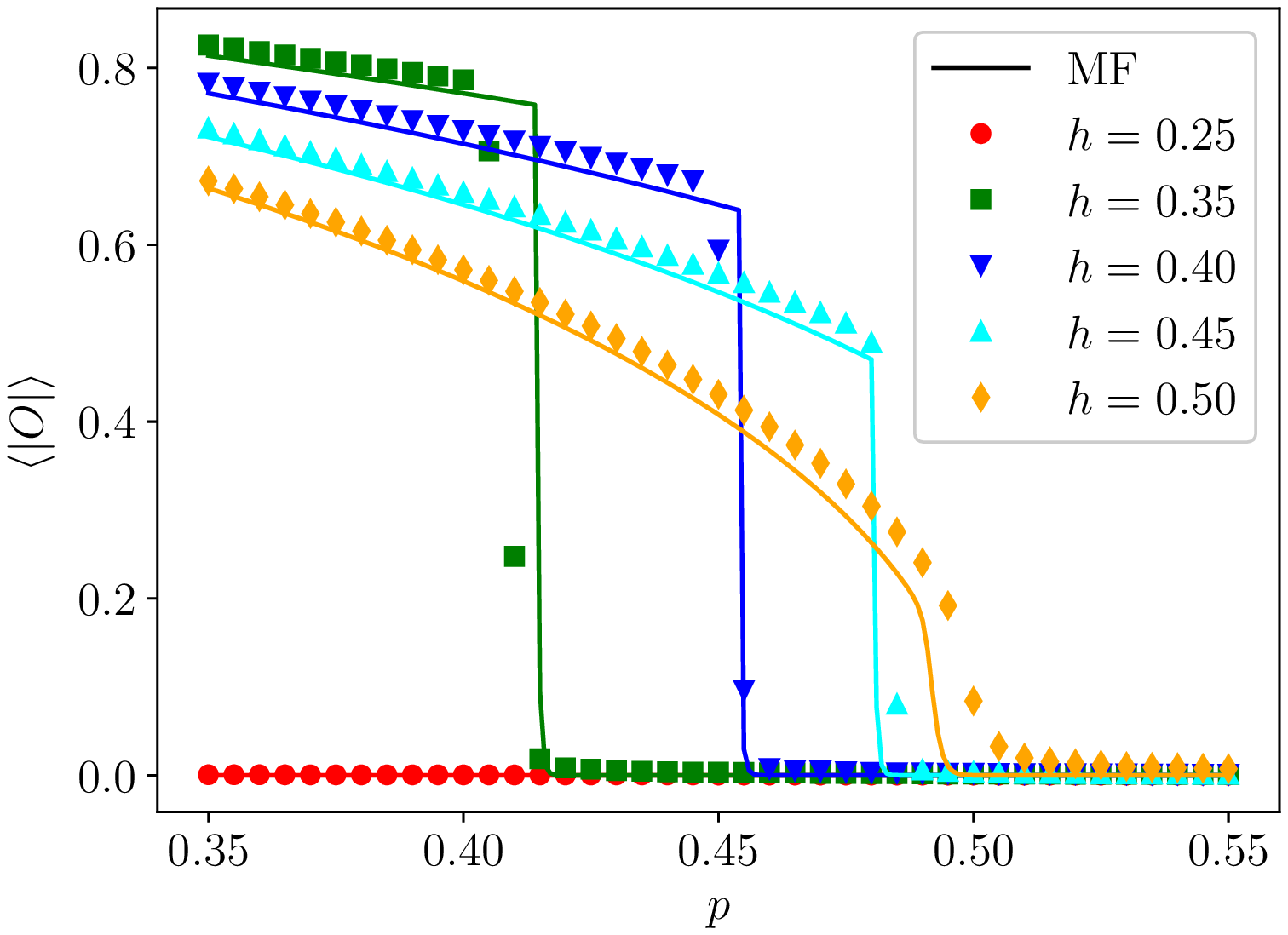}} \\
    \subfloat[]{\includegraphics[width=0.45\textwidth]{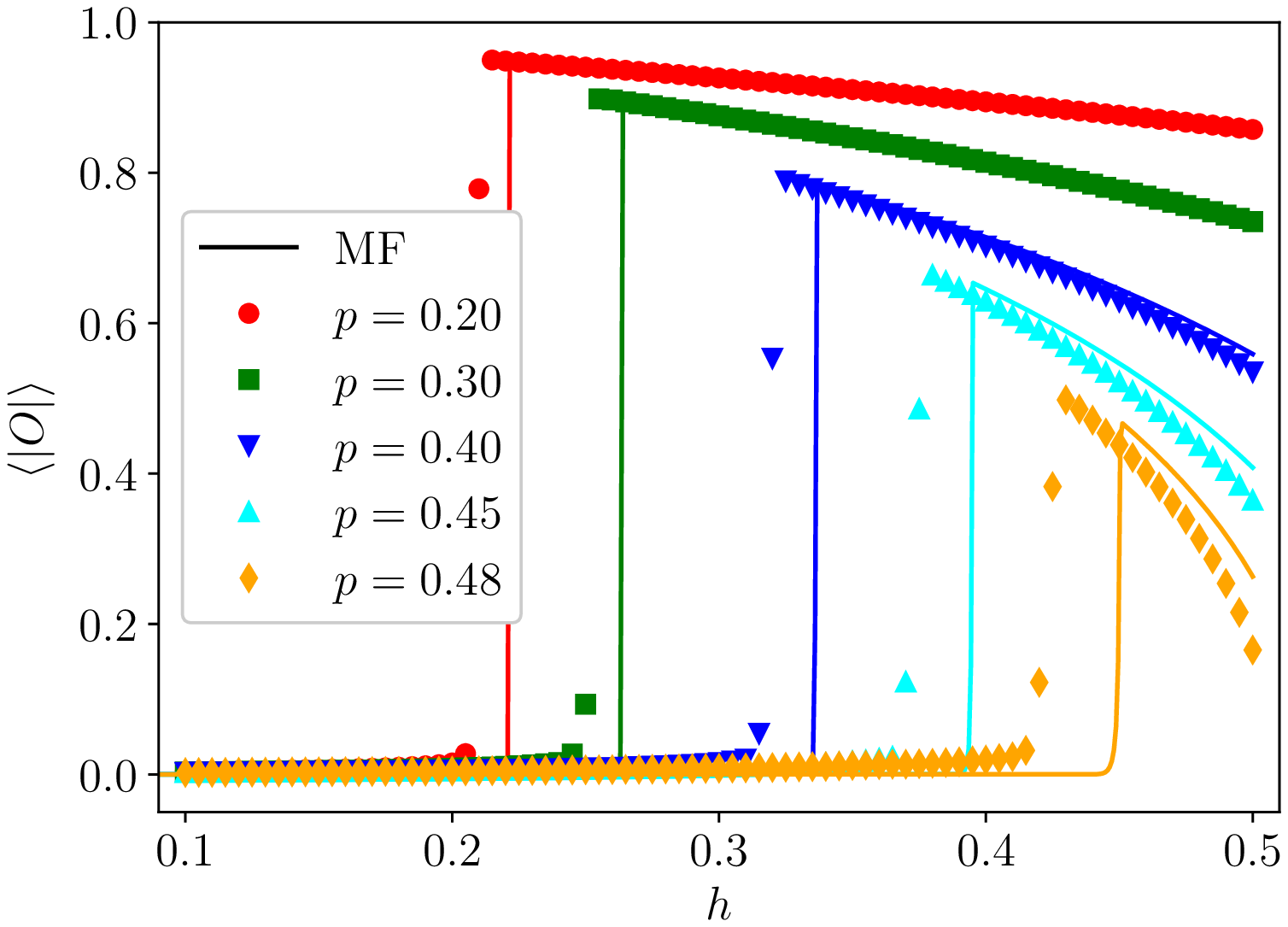}}
    \caption{Global order $O$ versus $p$ for some values of $h$ (a) and global order versus $h$ for some values of $p$ (b). Both graphs have $n_1=1/2$, $m_1 = 1$ and $m_2 =-1$ as initial conditions. The results act as a comparison between the approximated model and the model simulated in modular networks, here we see that they are in good agreement. The approximated solution was found via Euler integration of \cref{da,du,db} and the numerical simulations were performed with population size $N=10^4$ and averaged over 100 simulations in a network with $\<k>=30$. }
    \label{MD_EU_OxP}
\end{figure}

\begin{figure}[h]
    \centering
    \subfloat[]{\label{MD_EU_OxP_a}\includegraphics[width=0.45\textwidth]{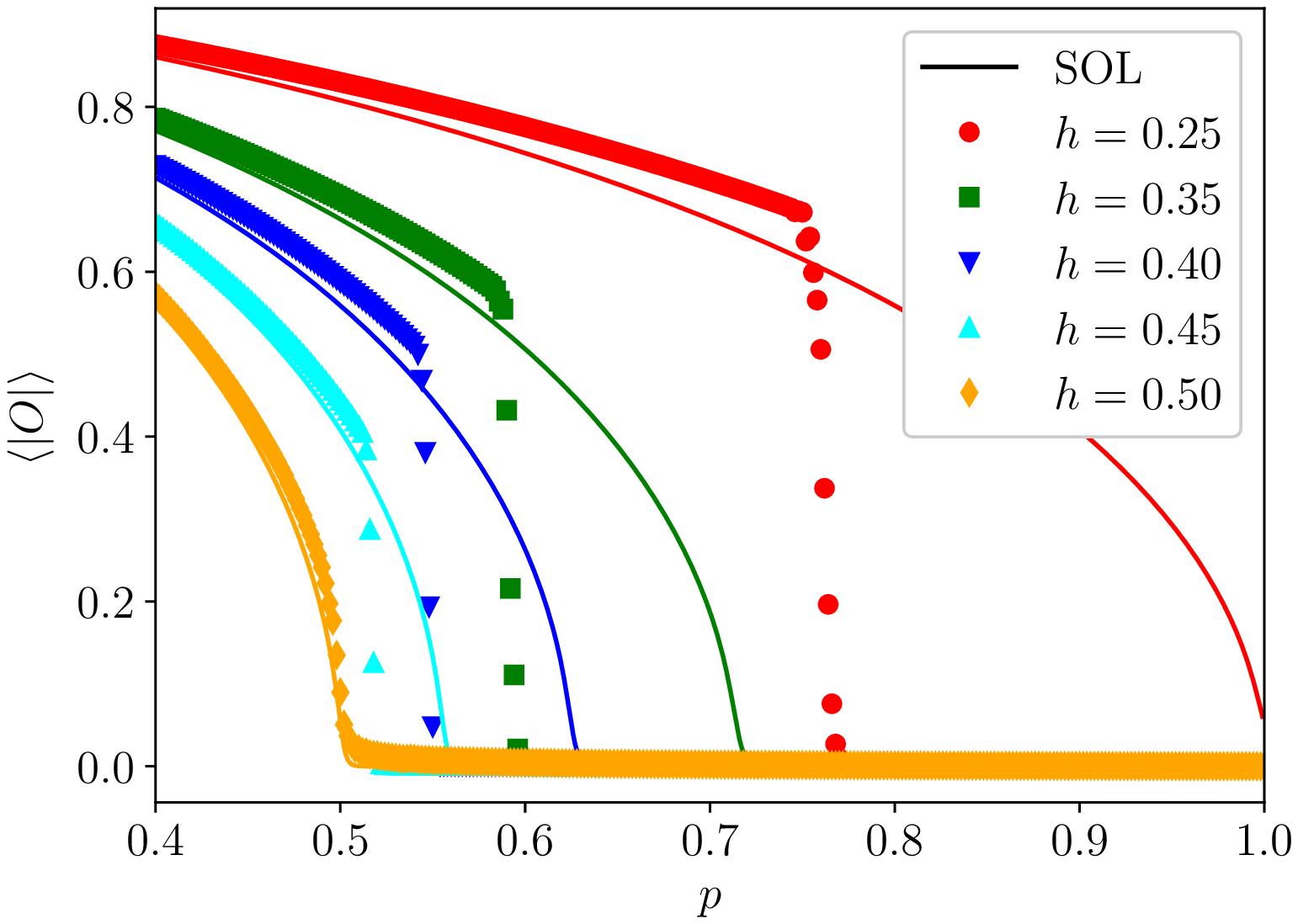}} \\
    \subfloat[]{\label{MD_EU_OxP_b}\includegraphics[width=0.45\textwidth]{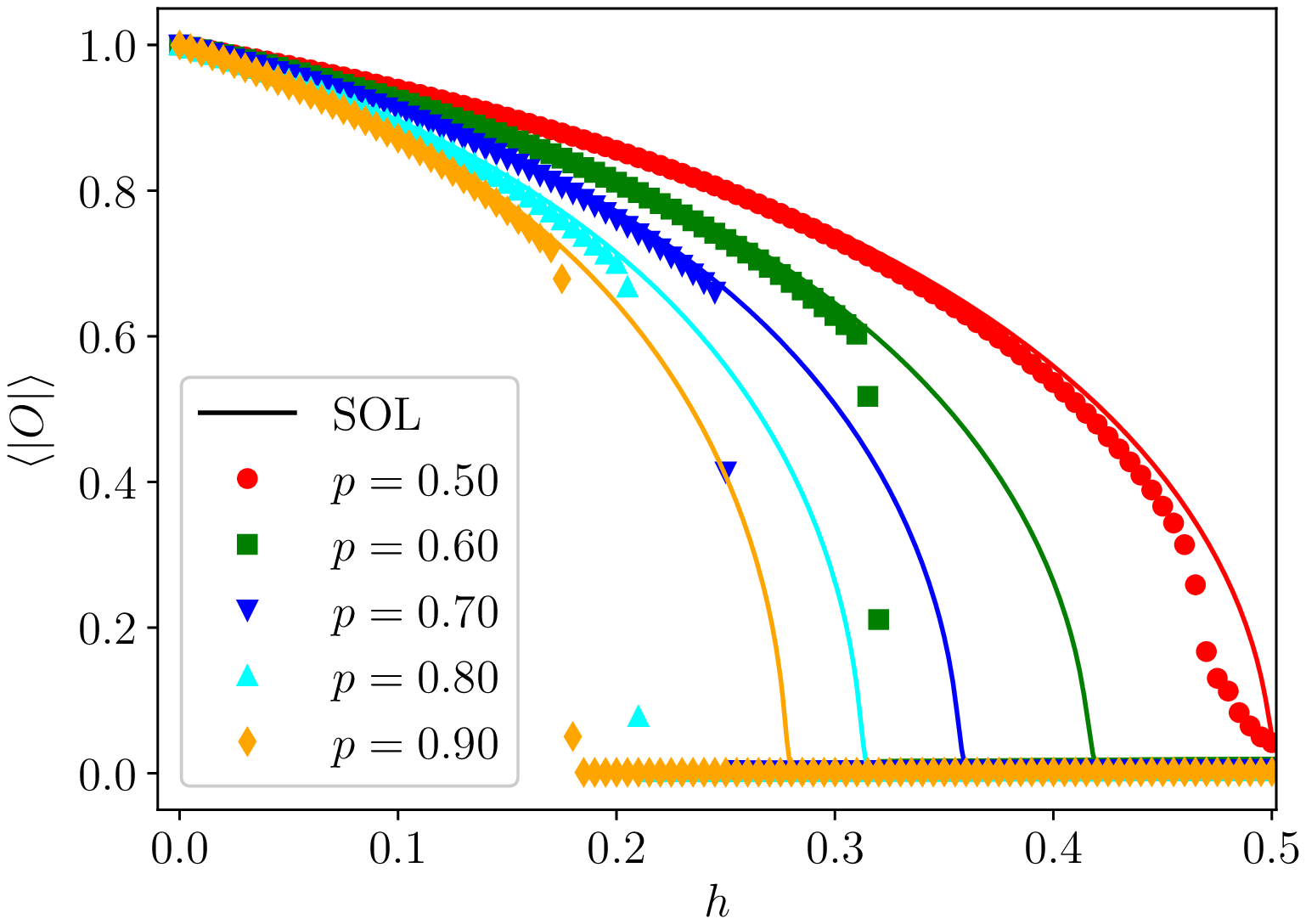}}
    \caption{Order parameter versus $p$ for some values of $h$ in (a), order parameter versus $h$ for some values of $p$ in (b). Both graphs have $n_1=1/2$ and $m_1 = m_2 = 1$ as initial conditions. The results act as a comparison between the approximated model and the model simulated in modular networks, here we see good agreement for values smaller than the critical point. The approximated solution (SOL) comes from \cref{MFsol} and the numerical simulations were performed with population size $N=10^4$ and averaged over 100 simulations in a network with $\<k>=30$. }
    \label{MD_EU_OxH}
\end{figure}

In \cref{MEnegativebias} we develop an analytic approach to better understand the behavior of our system. In this approach we consider that each community is fully connected like a mean-field approximation, but the individuals of a community can interact with a random individual of the other community with probability $P(i,j)$, as shown in \cref{zij}. Although this approximation ignores details of the network structure, it still mimics the community behavior of the system. The results obtained from our master equations and Monte Carlo simulations show good agreement, as can be seen in \cref{MD_EU_OxP}, except near the criticality when the order parameter of both communities start with same sign, as can be seen in \cref{MD_EU_OxH}. 

To facilitate the analysis we considered mainly communities of the same size ($n_1=n_2=1/2$). This scenario already encapsulates the significant results because these results come from the interactions between communities as cohesive units. In this case, we were also able to find the analytical curve that describes the ordered state of the system. 

In the stationary state with communities of same size ($n_1 = n_2 = 1/2$) the ordered state solution of the master equations for this model is given by (see \cref{MEnegativebias})

\begin{equation}
    \label{MFsol}
    O = \frac{\sqrt{1-4hp}}{1-hp}.
\end{equation}

\noindent This equation matches perfectly the numerical integration of \cref{da,du,db} when both communities start with $m_1 = m_2 = 1$, that can be seen in \cref{EU_OxHxP} (c). This result also describes very well the order parameter in the ordered state.

In \cref{EU_OxHxP} we exhibit the order parameter in the plane $h$ versus $p$ for distinct initial conditions. The results were obtained by numerical integration of the \cref{da,du,db}. The initial conditions of the graphics are $m_1=1$ and $m_2=-1$ (a), $m_1=0.02$, $u_1=0$, $m_2=0.0$ and $u_2 = 0.01$ (b), $m_1=m_2=1$ (c); $m_1=0.04$, $u_1=0.33$, $m_2=0.02$ and $u_2 = 0.33$ (d). One can see that the phase transition can be discontinuous for some values of the parameters. For example, in \cref{EU_OxHxP} (a) the order parameter $O$ drops from $O=1$ to $O=0$ when we increase $p$ for small values of $h$, when the network presents a clear community structure (see \cref{netws}). However, in many cases we see that the order parameter goes continuously from $1$ to $0$, as was predicted analytically in \cref{MFsol}. \Cref{EU_OxHxP} (d) shows an unusual behavior that can only be found for a very specific set of initial conditions, this indicates the presence of metastability in the system.

As we can see in \cref{MD_EU_OxP} the results for the approximated model are very similar to the results for the Monte Carlo simulations on the modular network. This is specially true when the communities start in disagreement, i.e.\ the order parameters of the communities start with different signs. The numerical integration only fails to reproduce the discontinuous phase transition when the communities start in agreement, i.e.\ the order parameters in both communities start with the same sign, as can be seen in \cref{MD_EU_OxH}.

This disagreement seems to steam from the finite size fluctuations of the system. The model has two metastable solutions. One in which the communities are aligned symmetrically and another in which they are aligned anti-symmetrically, these are described in more details in \cref{MEnegativebias}. In the mean field approach there are no system fluctuations, so we do not see the sudden transition from one metastable solution to another.

One can observe discontinuous phase transitions for some values of the parameters in the graphics of \cref{MD_EU_OxP}. These discontinuous phase transitions rise from the alignment of the communities. In the stationary state, communities can only align either symmetrically or anti-symmetrically. The discontinuous phase transition occurs when the system goes from the symmetrical to the anti-symmetrical arrangement.

The communities flipping as a whole instead of the individuals progressively flipping might be introducing inertia to the opinion changes. This happens because the communities only interact positively, therefore promoting local consensus. This result is in line with \cite{2018encinasHOF}, where the authors found that opinion inertia gives rise to a discontinuous phase transition in the majority-vote model.

The \cref{MD_EU_OxP} (b) shows an interesting nonmonotonic ordering: the increase of order for raising $h$, and a subsequent decrease of the order parameter for higher values of $h$. In order to better understand this unusual behavior one needs to keep in mind that combined with intergroup bias the consequences of intergroup connectivity are twofold. In one hand if the intergroup connectivity is too low there is no way for the opinions of one group to connect to the other group, and thus produce consensus. On the other hand higher intergroup connectivity also increases the probability of a negative interaction which reduces the global order parameter.

A similar nonmonotonic ordering was found in a continuous model of opinion dynamics \cite{2017anteneodoC} and also in a q-voter model with independence and memory  \cite{2018jedrzejewskiS}. Our work adds a novel mechanism for the emergence of nonmonotonic phenomena in social scenarios: the combination of community structure and negative intergroup interactions in three-state opinion dynamics.


\section{Model B: inflexibles and noise}

In this section, differently from the model presented in the previous section, all interactions are positive for simplicity. A fraction of the population does not change opinion (inflexible agents) and agents can take the neutral opinion due to noise.

\subsection{Description}

At the beginning of the simulation we generate the network, as discussed in \cref{genNet}. Then randomly pick a fraction $f$ of the total population to be inflexible. For the purposes of this model we have all inflexibles in community 1. The initial opinion of the inflexibles is set to $+1$ and the opinions of all other agents are set to $-1$.

At a given time step $t$ each agent $i$ that is not inflexible will update its opinion. With probability $q$ agent $i$ becomes neutral, i.e. $o_i = 0$ \cite{2013crokidakis,2005bennaim}. With complementary probability $1-q$ we choose one of its neighbors $j$ at random. Then update agent's $i$ opinion according to \cref{int_pos}.

For the model considered in this section, it is also possible to obtain master equations by means of a coupled mean-field approximation, as we discuss in \cref{MEinflex}.


\subsection{Results and discussion}

\begin{figure}[h]
    \centering
    \includegraphics[width=0.45\textwidth]{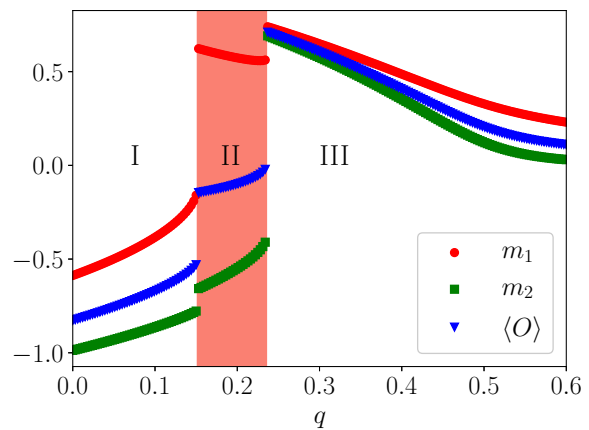}
    \caption{Numerical integration showing the layout of the different phases, for $n_1=0.4, h=0.1$ and $f=0.05$. In the first region (I) the opinion of the inflexible agents does not become the majority opinion in neither community. In the second region (II) the inflexibles' opinion $+1$ become the majority opinion in their community. And finally, in the third region (III) the opinion of inflexibles has become majority in both communities. Intermediate values of $q$ promote the spread of minority opinion, but values of $q$ that are too high end up weakening the spread of the minority opinion.}
     \label{IN_phases}
\end{figure}

In \cref{IN_phases} we exhibit the order parameter $O$ as a function of the noise $q$ for a fraction $f=0.05$ of inflexibles, as well as the local order parameters $m_{1}$ and $m_{2}$, as defined in \cref{eq_m_i}. The data were obtained by the numerical integration of \cref{ba1,ba2,bb1,bb2,bu1,bu2}. The community 1 has a relative size $n_1=0.4$, i.e. the inflexible agents are located in the smaller community. We are interested in verifying if the opinion of a minority fraction of the population (the inflexibles) can become the majority opinion locally in community 1, as well as the global majority opinion in both communities. \Cref{IN_phases} shows 3 regions, labeled by I, II and III\@. In region I, for $q\lessapprox 0.15$, the opinion of the inflexibles does not spread over the network, and the opinion $o=+1$ remains the minority opinion even in community 1. In region II, the opinion of the inflexible agents will be shared by the majority of agents in community 1. Finally, in region III the inflexible initial minority opinion spreads fast and it becomes the majority opinion in all the network, i.e. in both communities 1 and 2. It is interesting to observe such minority reversion even for a very small fraction of inflexibles, around $5\%$. Such kind of minority reversion was observed before in simple opinion dynamics models \cite{2002galam,2008huangCWQ,2009huangCQ,2010shenL,2017wuXZ,2014crokidakisO}, but to the best of our knowledge it is the first time that it is due to the presence of inflexibility in the population.

It is important to observe that the minority opinion (opinion of the inflexibles) only spreads over the network and becomes the majority if the neutrality noise is present. We verified that the presence of inflexibles in the model with intergroup bias does not lead to global takeover by the inflexibles.

\begin{figure*}[t]
    \centering
    \subfloat[]{\includegraphics[width=0.4\textwidth]{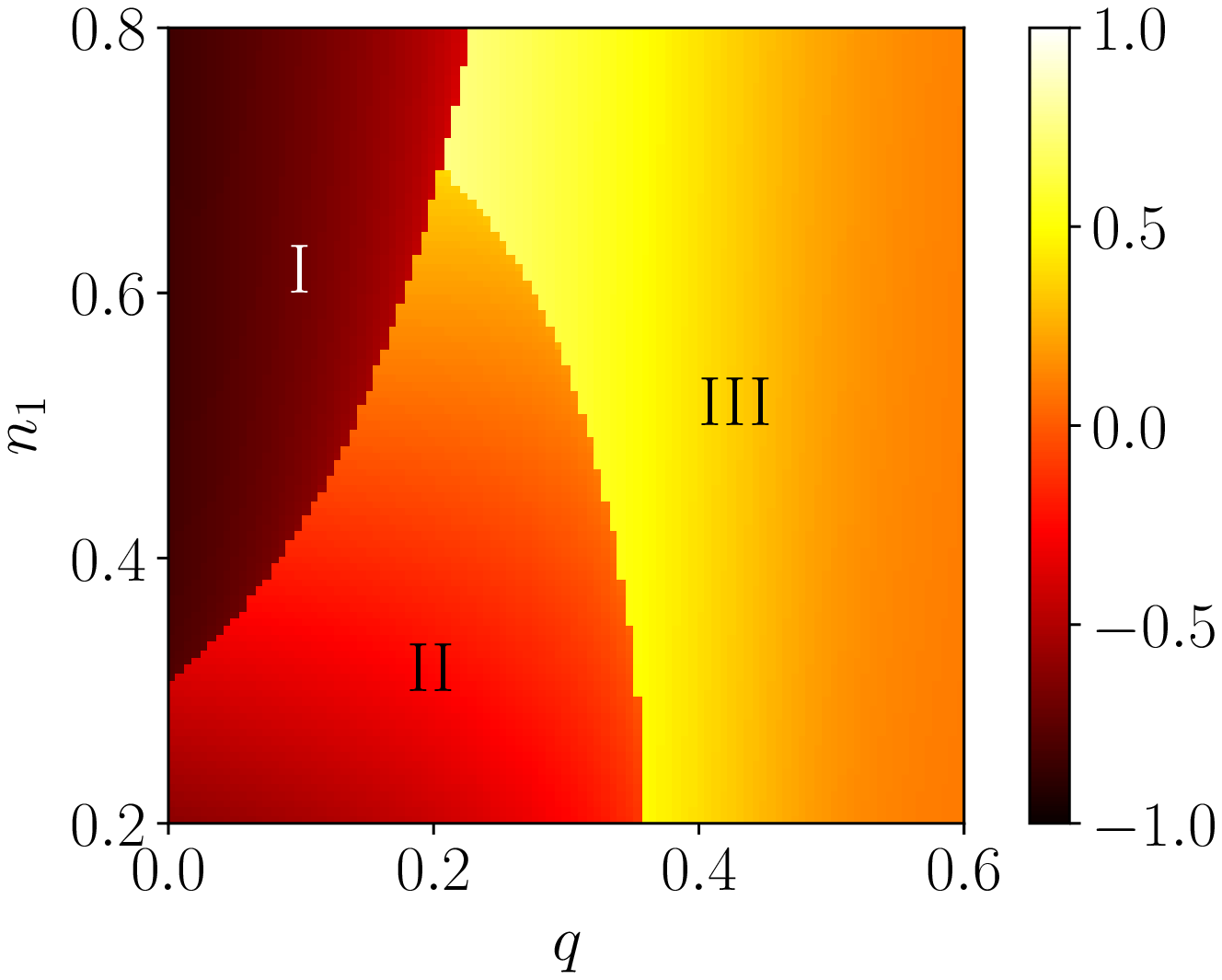}}
    \subfloat[]{\includegraphics[width=0.4\textwidth]{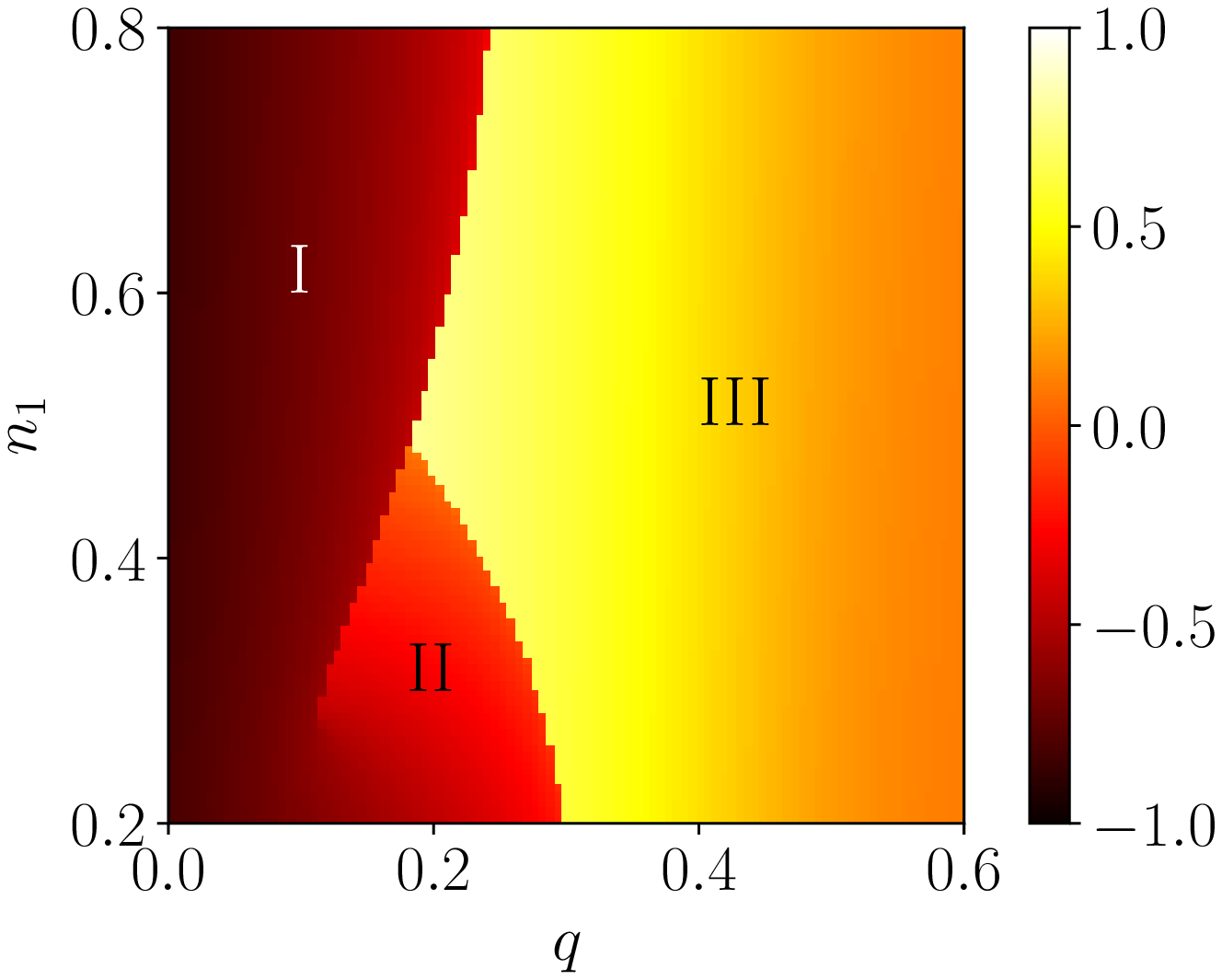}}
    \\
    \subfloat[]{\includegraphics[width=0.4\textwidth]{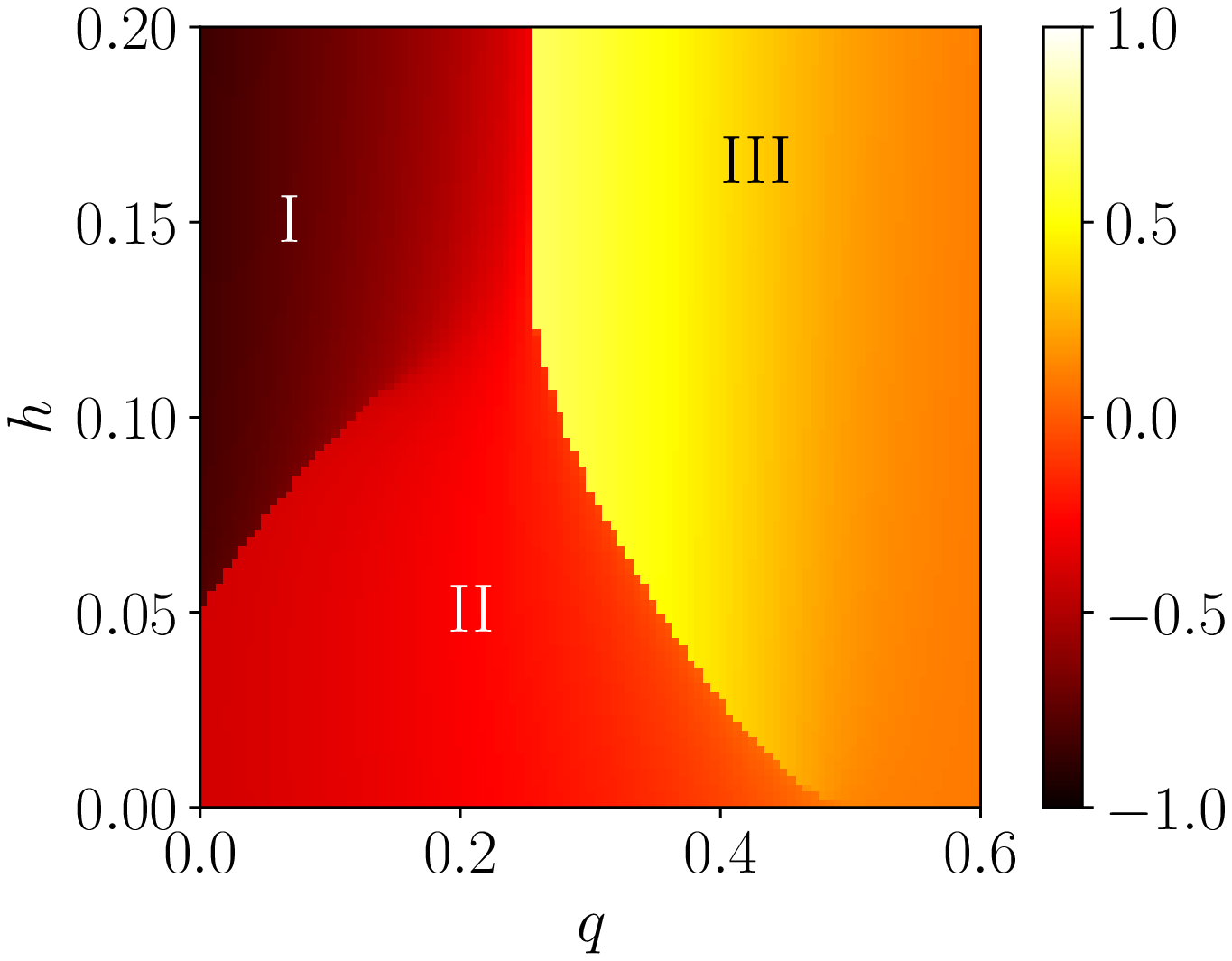}} 
    \subfloat[]{\includegraphics[width=0.4\textwidth]{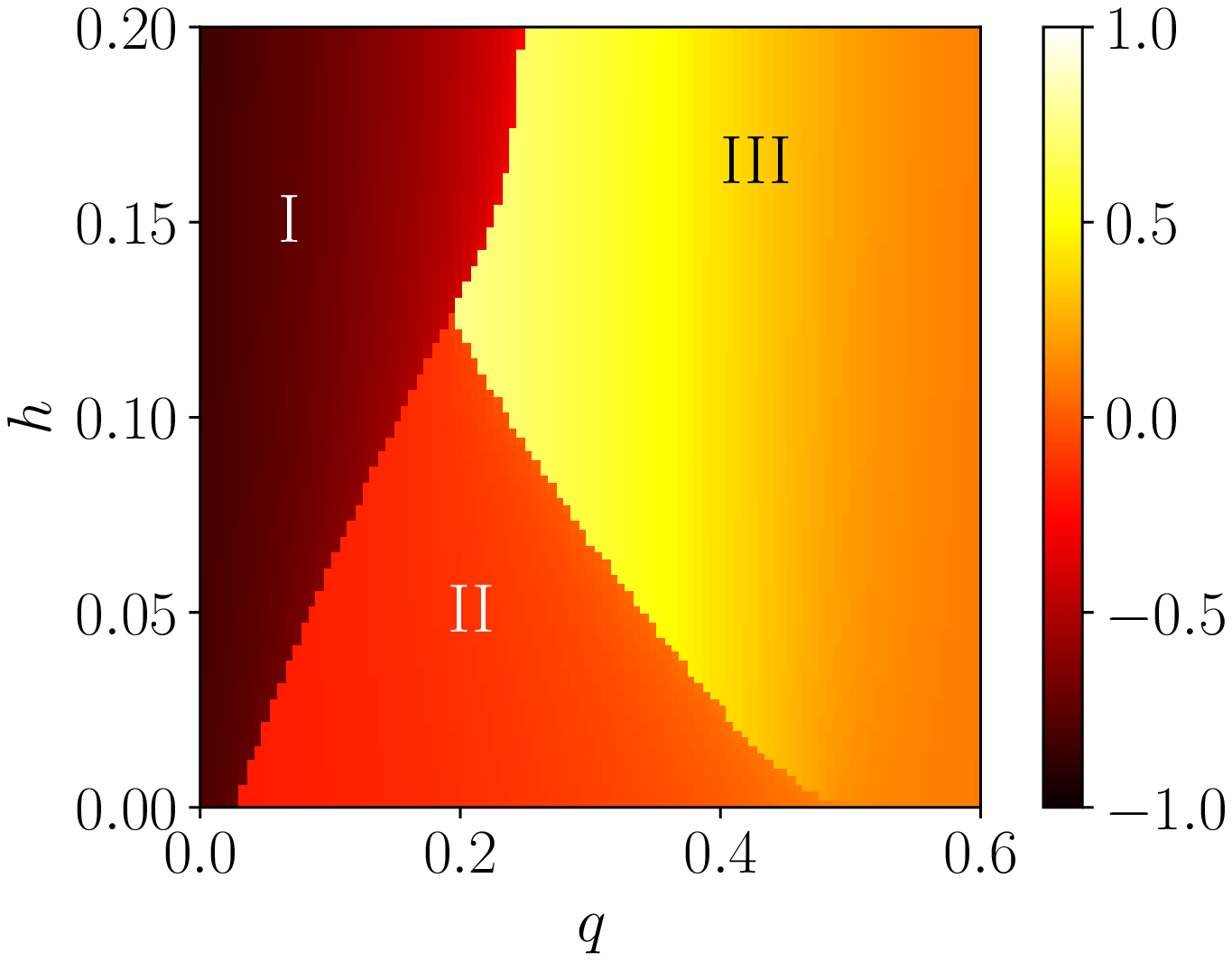}}
    \\
    \subfloat[]{\includegraphics[width=0.4\textwidth]{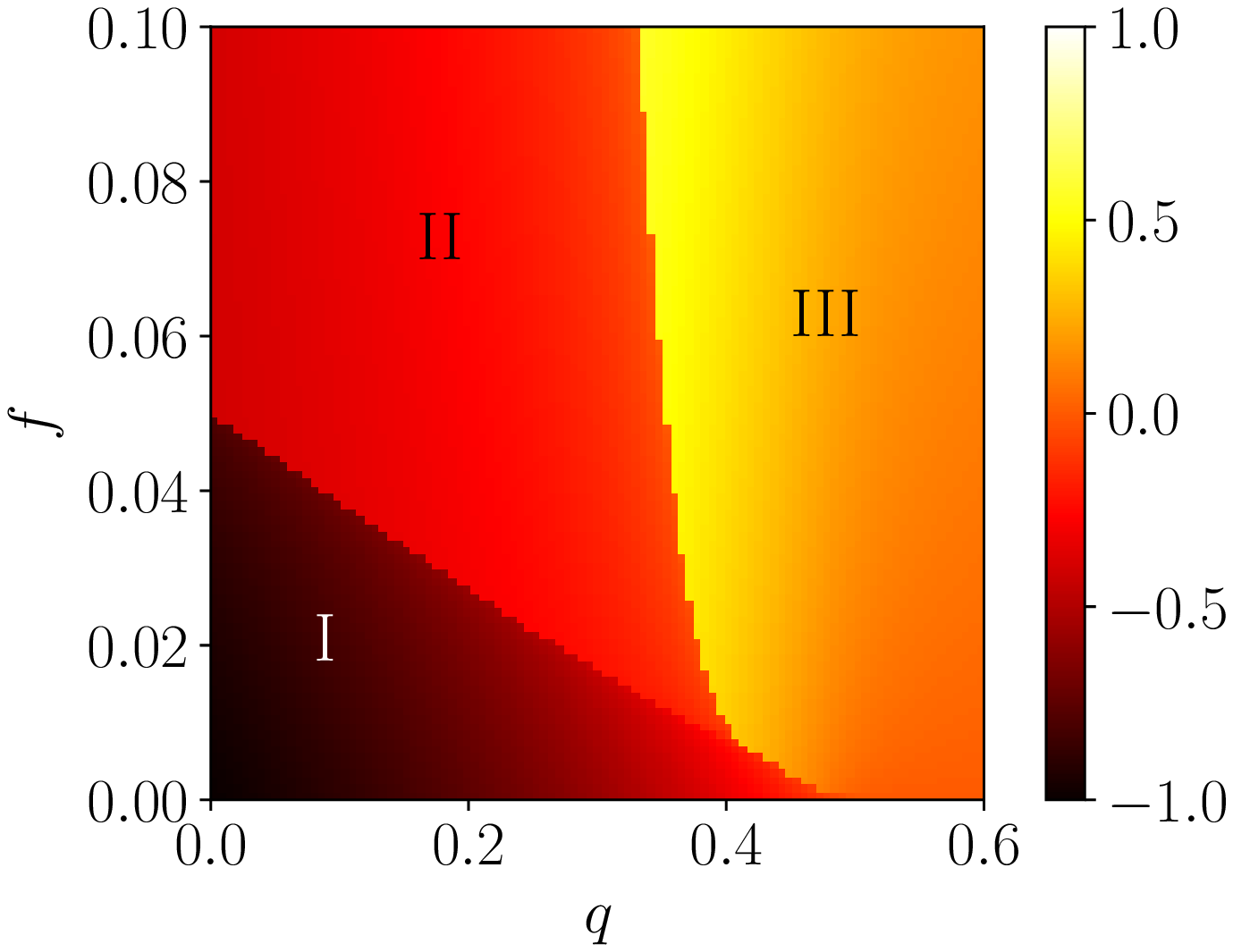}}
    \subfloat[]{\includegraphics[width=0.4\textwidth]{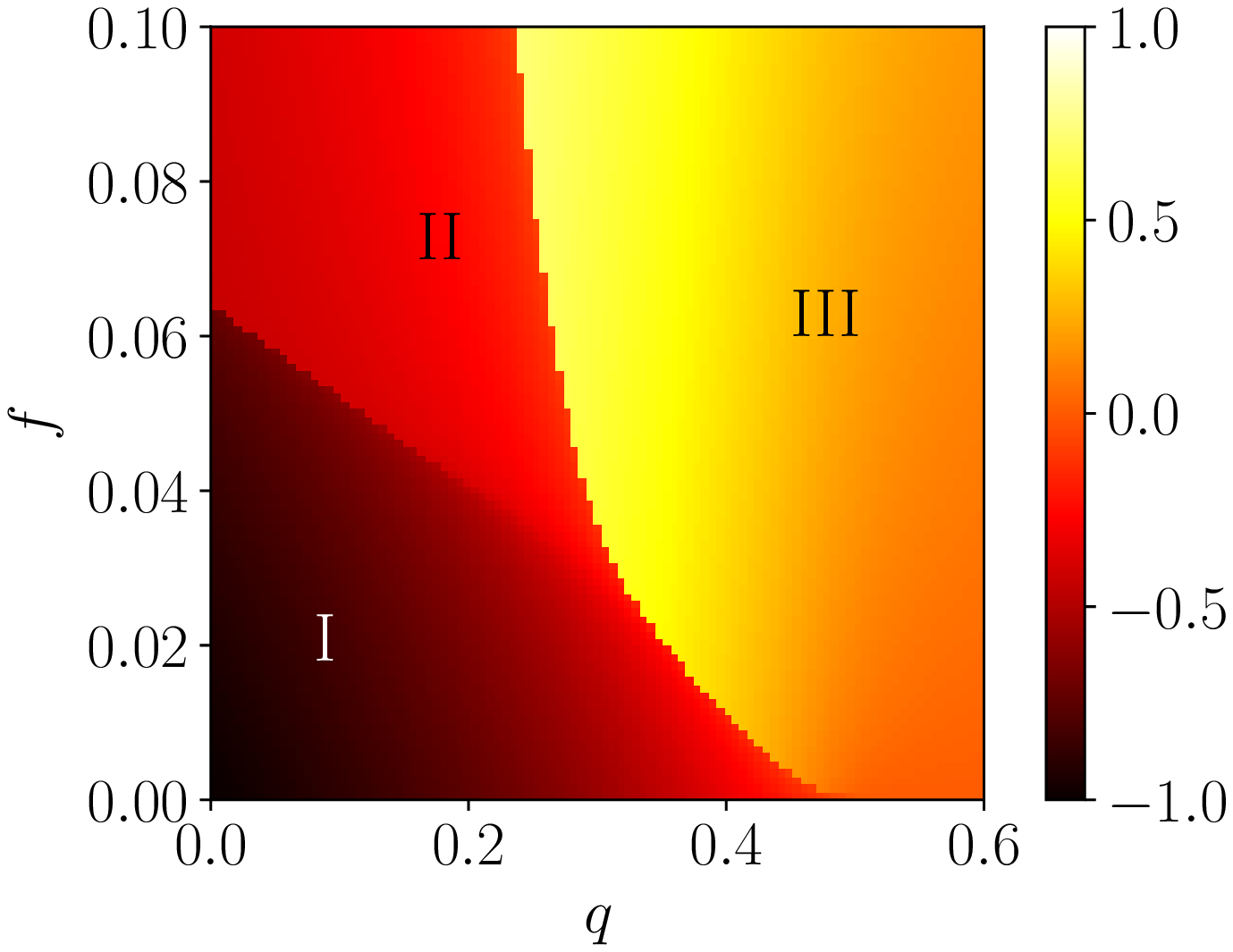}}
    \caption{Global order parameter $O$  for various configurations arising from the combinations of parameters. Specifically $n_1$ vs $q$ with $h=0.05 \; f = 0.05$  in (a),$n_1$ vs $q$ with $h=0.10 \; f = 0.05$  in (b), $h$ vs $q$ $n_1=0.30 \; f = 0.05$ in (c), $h$ vs $q$ $n_1=0.40 \; f = 0.05$ in (d),  $f$ vs $q$ with $n_1=0.30 \; h = 0.05$ in (e) and $f$ vs $q$ with $n_1=0.30 \; h = 0.10$ in (f).}
    \label{IN_EU}
\end{figure*}

In \cref{IN_EU} we exhibit phase diagrams of the model in distinct planes, namely $n_1$ versus $q$ (panels a and b), $h$ versus $q$ (panels c and d) and $f$ versus $q$ (panels e and f). In the graphics one can see that the regions of local majority (region II, opinion $o=1$ of the inflexibles becomes majority in community 1) and global majority (region III, opinion $o=1$ becomes the majority in both communities) can be obtained for a wide range of the parameters. Indeed, the region III results in a competition of the parameters. For example, let us consider the region of weak noise $q$. Panels (a) and (b) show that the increase of the community with the presence of inflexibles (community 1) makes hard the spread of the opinion $o=1$. The increase of out-group interactions (raising $h$) decreases considerably region II. The decrease of the relative size of community 1, $n_1$, from panel (d) to (c), helps to spread the inflexible opinion $o=1$. Finally, the increase the fraction $f$ of inflexibles obviously leads to an increase of regions II and III, as one can see in panels (e) and (f). In all scenarios, the highest value of the global order parameter  $O$ (brightest region) occurs when the propensity to neutrality achieves moderated values meaning that the nonmonotonic global ordering with the noise strength $q$ is robust.

\begin{figure}[h]
    \centering
    \subfloat[]{\includegraphics[width=0.49\textwidth]{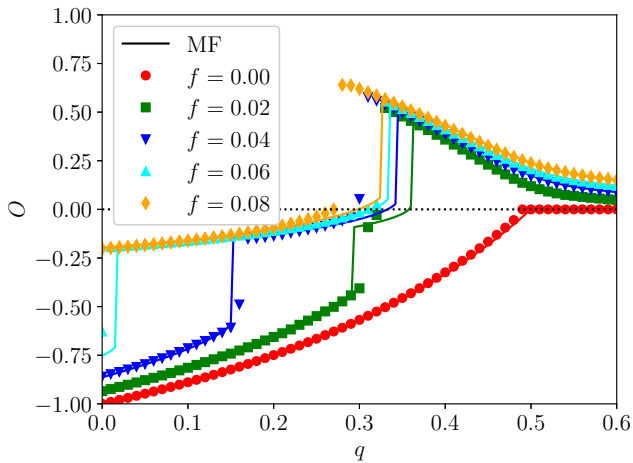}} \\
    \subfloat[]{\includegraphics[width=0.49\textwidth]{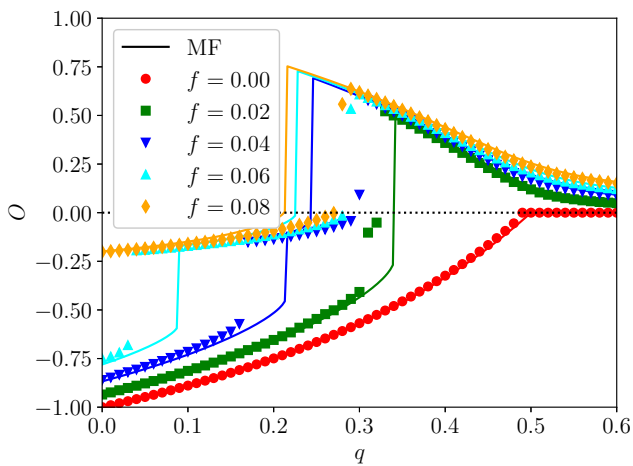}}
    \caption{Comparison between Monte Carlo simulations and numerical integration for the order parameter vs $q$ with $N=10^4$ and for $h=0.05 \; n_1 = 0.4$ in (a) and $h=0.10 \; n_1 = 0.4$ in (b). Here we can see that the approximated model has good predictable power except near the criticalities. It is curious that for the 1st phase transition it over predicts the critical point and for the second it under predicts it.}
    \label{IN_MC_EU}
\end{figure}

In \cref{IN_MC_EU} we compare the results of the numerical integrations of the equations of the model (from \cref{MEinflex}) and Monte Carlo simulations. One can see that, apart from the region next to the transitions, the master equations can capture the essence of the dynamics of the model.

\section{Final Remarks}
\label{conclusion}

In this work we study the opinion evolution in an artificial community-based population. The social network of contacts is represented by a modular network that presents a community structure.  We consider a parameter $h$ that controls the strength of the community structure: a large value of $h$ yields more links between the two communities and, thus, a weak community structure. We employ this modular network to address two questions. 

In the first problem there is another parameter $p$ that introduces disorder in the interactions, that can be positive or negative with probabilities $1-p$ and $p$, respectively. We study the model by means of analytical and numerical calculations. We found that the system exhibits order-disorder transitions, and for some values of the parameters $h$ and $p$ such transition can be discontinuous. In addition, we also found a disorder-induced transition for increasing h for a wide range of values of the disorder parameter p. This is not a usual result in models of opinion dynamics, but it was recently observed in a model of continuous opinions \cite{2017anteneodoC} and for a $q$-voter model \cite{2018jedrzejewskiS}. Our results also show that the introduction of intergroup bias is capable of promoting the polarization of opinions. The polarization can be observed by the anti-symmetrical alignment of the order parameters of the two communities.
This is in accordance with previous findings that political discussions over Twitter are both polarized and partisan \cite{2011conoverRFG}. Moreover, our results suggest that the intergroup bias is driving polarization, as was suggested in \cite{2010yardiB}.

In the second part, we considered another formulation of the opinion model, taking into account noise towards neutrality and an inflexible minority localized in one community. Our results show an interesting nonmonotonic global ordering when the strength of the noise is increased. That is the propensity to neutrality acts a double-edged sword: an intermediate intensity of the bias to neutrality is beneficial to the initial minority opinion spreads over the network, but this noise-assisted minority spreading is weakened if the neutrality is excessively favored in the population. This global reversal of opinion occurs abruptly.

In a recent work \cite{2018piresOC} it was discussed how abrupt changes in the global opinion of a population can affect the spreading of diseases when a vaccination campaign is taken into account. In the mentioned model, the opinions against and in favor of the vaccination influences directly the vaccination probability of the agents. As the modular structures we consider here lead to discontinuous transitions and nonmonotonic phenomena in both formulations of our model, it can be interesting to consider those structures to simulate the spreading of diseases taking into account the coupling of opinions and vaccination probability. This study will be considered in a future work.

\section*{Acknowledgments}

The authors thank Serge Galam for fruitful discussions. Financial support from the Brazilian funding agencies Conselho Nacional de Desenvolvimento Cient\'ifico e Tecnol\'ogico (CNPq), Coordena\c{c}\~ao de Aperfei\c{c}oamento de Pessoal de N\'ivel Superior (CAPES) and Funda\c{c}\~ao Carlos Chagas Filho de Amparo \`a Pesquisa do Estado do Rio de Janeiro (FAPERJ) is also acknowledged.



\appendix

\section{Master equations for model with intergroup bias}
\label{MEnegativebias}

We consider that each community is fully connected like a mean-field approximation, but the individuals of a community can interact with a random individual of the other community with probability $h$. 
In this approximation one can obtain the master equations of the system,

\begin{equation}
    \label{da}
    \begin{split}
        \dot{a_i} = u_i \left\{ \left( 1-\frac{h}{2n_i}\right)a_i + \frac{h}{2n_i}\left[ (1-p)a_j + pb_j \right] \right\} \\
        - a_i \left\{ \left(1 - \frac{h}{2n_i} \right)b_i +\frac{h}{2n_i}[pa_j + (1-p)b_j]\right\},
    \end{split}
\end{equation}

\begin{equation}
    \label{du}
    \begin{split}
        \dot{u_i} = a_i\left\{ \left(1-\frac{h}{2n_i} \right)b_i +\frac{h}{2n_i}[pa_j + (1-p)b_j] \right\} \\ + b_i\left\{ \left(1 - \frac{h}{2n_i} \right)a_i +\frac{h}{2n_i}[pb_j + (1-p)a_j]\right\} \\ - u_i\left\{ \left(1 - \frac{h}{2n_i} \right)(a_i+b_i) + \frac{h}{2n_i}[a_j+b_j]\right\} ,
    \end{split}
\end{equation}

\begin{equation}
    \label{db}
    \begin{split}
    \dot{b_i} = u_i\left\{ \left(1 - \frac{h}{2n_i} \right)b_i + \frac{h}{2n_i}\left[ (1-p)b_j + pa_j\right]\right\} \\
        - b_i\left\{ \left(1-\frac{h}{2n_i} \right)a_i +\frac{h}{2n_i}[pb_j + (1-p)a_j]\right\}.
    \end{split}
\end{equation}

\noindent In above equations $i,j=1,2$ with $i\neq j$, $a_i$ is the density of negative opinions ($o = -1$), $u_i$ is the density of neutral opinions ($o = 0$) and $b_i$ is the density of positive opinions ($o = +1$) in the community $i$.

These equations were numerically integrated using the Euler method, considering a step size $dt=0.1$ and a maximum time $t_{max} = 10000$. In \cref{MD_EU_OxP} we see a good agreement between the numerical integration of the above master equations and our Monte Carlo simulations.

Considering communities of the same size ($n_1 = n_2 = 1/2$) we can obtain a steady-state solution by means of an ansatz. A preliminary inspection of the time series insightfully reveals two main types of steady-state solutions:

 \begin{itemize}
\item (I)  $a_i^\infty=b_j^\infty$,
$b_i^\infty=a_j^\infty$,
$u_i^\infty=u_j^\infty$

\item (II)  $a_i^\infty=a_j^\infty$,
$b_i^\infty=b_j^\infty$,
$u_i^\infty=u_j^\infty$
  \end{itemize}

\noindent In a nutshell, this means that in the steady state the communities can be either anti-symmetrically (I) or symmetrically (II)  aligned. A more mathematically inclined reader can also see that \cref{da,db,du} possess these symmetries when the communities have the same size.

The ansatz for the case I leads to the disordered phase
\begin{align}
 O^\infty = 0 \quad\quad \text{I: disordered solution}
 \label{disorder}
\end{align}
On the other hand, the insertion of the ansatz for the case II into \cref{da,db} gives

\begin{equation}
    \label{da2}
 (1-hp)u_1^\infty  a_1^\infty  + hpu_1^\infty b_1^\infty  - (1-hp)a_1^\infty b_1^\infty  - hp(a_1^\infty )^2 = 0
\end{equation}
\begin{equation}
    \label{db2}
 (1-hp)u_1^\infty b_1^\infty  + hpu_1^\infty a_1^\infty  - (1-hp)a_1^\infty b_1^\infty  - hp(b_1^\infty )^2 =  0
\end{equation}

Subtracting \cref{da2} from \cref{db2} gives the trivial solution $ a_1^\infty = b_1^\infty $ ( $ O^\infty = 0 $) and the steady-state fraction of undecided agents

\begin{equation}
    \label{eqintermediario2}
u_1^\infty  =  \frac{hp}{1-hp}
\end{equation}
where we have used  $a_1^\infty + b_1^\infty + u_1^\infty = 1$.

From \cref{eqintermediario2} and \cref{da2} we obtain

\begin{equation}
    \label{da3}
    (a_1^\infty)^2 - \left( \frac{1-2hp}{1-hp} \right)a_1^\infty + \left( \frac{hp}{1-hp}  \right)^2  = 0
\end{equation}

Then

\begin{equation}
    \label{da4}
    a_1^\infty  =  \frac{1}{2(1-hp)}
    \left( 1-2hp \pm \sqrt{1-4hp }  \right)
\end{equation}

From $ O^\infty = | \frac{(a_1^\infty+a_2^\infty)-(b_1^\infty+b_2^\infty)}{2} | = |a_1^\infty - b_1^\infty |= |a_1^\infty - (1-u_1^\infty-a_1^\infty) |$ and
 \cref{eqintermediario2,da4} we finally get

\begin{align}
 O^\infty  = \frac{\sqrt{1-4hp} }{1-hp} \quad\quad \text{II: ordered solution.}
 \label{order}
\end{align}

\Cref{disorder,order} show the presence of an order-disorder phase transition in our dynamics,  but these equations do not show explicitly the discontinuous-continuous boundary that we have observed in the main part of the manuscript. This seems to happen because the discontinuous phase transition rises from the change of ansatz. Despite this, there is a reasonable agreement between the \cref{order} and the Monte Carlo simulations for a large set of parameters, as shown in \cref{MD_EU_OxH}.


\section{Master equations for model with inflexibles and noise}
\label{MEinflex}

Let us first turn our attention to the model with the noise that makes agent's opinions neutral, without considering the inflexibles. This makes the problem easier to solve due to the still present symmetry. Without the inflexibles the normalization rule is $a_i+b_i+u_i=1$. In the infinite population limit we have:

\begin{equation}
    \begin{split}
    \dot{a_i} = (1-q) \left\{ u_i \left[ \left( 1- \frac{h}{2n_i} \right) a_i + \frac{h}{2n_i}a_j \right]\right. \\
        \left. - a_i \left[ b_i\left( 1 - \frac{h}{2n_i} \right) + \frac{h}{2n_i}b_j \right] \right\} -q a_i
    \end{split}
\end{equation}

\begin{equation}
    \begin{split}
        \dot{u_i} = q(a_i + b_i) + (1-q)\left\{ a_i\left[ \left( 1-\frac{h}{2n_i} \right) b_i +\frac{h}{2n_i}b_j \right] \right. \\ 
        + b_i\left[ \left( 1-\frac{h}{2n_i} \right) a_i +\frac{h}{2n_i}a_j \right] \\
        \left.-u_i \left[ (a_i +b_i)\left( 1- \frac{h}{2n_i} \right) + \frac{h}{2n_i} (a_j +b_j) \right] \right\}
    \end{split}
    \label{dui}
\end{equation}

\begin{equation}
    \begin{split}
    \dot{b_i} = (1-q) \left\{ u_i \left[ \left( 1- \frac{h}{2n_i} \right) b_i + \frac{h}{2n_i}b_j \right]\right. \\
        \left.- b_i \left[ a_i\left( 1- \frac{h}{2n_i} \right) + \frac{h}{2n_i}a_j \right] \right\} -q b_i
    \end{split}
\end{equation}

Now if we introduce a fraction $f$ of inflexibles all in the community 1 the normalization rules become $a_1+b_1+u_1+f/n_1 = 1$ and $a_2+b_2+u_2 = 1$. Notice that the fraction of inflexibles is limited by the relative size of community 1 ($n_1$). In the infinite population limit we get

\begin{equation}
    \begin{split}
        \dot{a_1} = (1-q) \left\{ u_1 \left[ \left( 1- \frac{h}{2n_1} \right) a_1 + \frac{h}{2n_1}a_2  \right] \right. \\
        \left. - a_1 \left[ \left(b_1 + \frac{f}{n_1} \right)\left( 1 - \frac{h}{2n_1} \right) + \frac{h}{2n_1}b_2 \right] \right\} -q a_1
    \end{split}
    \label{ba1}
\end{equation}

\begin{equation}
    \begin{split}
        \dot{u_1} = q(a_1 + b_1) \\
        +(1-q)\left\{ a_1\left[ \left( 1-\frac{h}{2n_1} \right) \left(b_1 + \frac{f}{n_1} \right) +\frac{h}{2n_1}b_2 \right] \right. \\ 
        + b_1\left[ \left( 1-\frac{h}{2n_1} \right) a_1 +\frac{h}{2n_1}a_2 \right] \\
        \left. -u_1 \left[ (1 - u_1)\left( 1- \frac{h}{2n_1} \right) + \frac{h}{2n_1} (a_2 +b_2) \right] \right\}
    \end{split}
    \label{bu1}
\end{equation}

\begin{equation}
    \begin{split}
        \dot{b_1} = (1-q) \left\{ u_1 \left[ \left( 1- \frac{h}{2n_1} \right) \left(b_1 + \frac{f}{n_1} \right) + \frac{h}{2n_1}b_2 \right] \right. \\
        \left. - b_1 \left[ a_1\left( 1- \frac{h}{2n_1} \right) + \frac{h}{2n_1}a_2 \right] \right\} -q b_1
    \end{split}
    \label{bb1}
\end{equation}

\begin{equation}
    \begin{split}
        \dot{a_2} = (1-q) \left\{ u_2 \left[ \left( 1- \frac{h}{2n_2} \right) a_2 + \frac{h}{2n_2}a_1 \right]  \right. \\
        \left. - a_2 \left[ b_2\left( 1 - \frac{h}{2n_2} \right) + \frac{h}{2n_2}\left(b_1+\frac{f}{n_1} \right) \right] \right\} -q a_2
    \end{split}
    \label{ba2}
\end{equation}

\begin{equation}
    \begin{split}
        \dot{u_2} = q(a_2 + b_2) \\
        + (1-q)\left\{ a_2\left[ \left( 1-\frac{h}{2n_2} \right) b_2 +\frac{h}{2n_2}\left(b_1 + \frac{f}{n_1} \right)\right] \right. \\ 
       + b_2\left[ \left( 1-\frac{h}{2n_2} \right) a_2 +\frac{h}{2n_2}a_1 \right]  \\
         \left. -u_2 \left[ (a_2 +b_2)\left( 1- \frac{h}{2n_2} \right) + \frac{h}{2n_2} (1 -u_1) \right] \right\}
    \end{split}
    \label{bu2}
\end{equation}

\begin{equation}
    \begin{split}
        \dot{b_2} = (1-q) \left\{ u_2 \left[ \left( 1- \frac{h}{2n_2} \right) b_2 + \frac{h}{2n_2}\left(b_1 + \frac{f}{n_1} \right) \right]  \right. \\
         \left. - b_2 \left[ a_2\left( 1- \frac{h}{2n_2} \right) + \frac{h}{2n_2}a_1 \right] \right\} -q b_2
    \end{split}
    \label{bb2}
\end{equation}

In this case it was not possible to obtain an explicit solution for the steady-state. But in \cref{IN_MC_EU} we see a good agreement between our Monte Carlo simulations and numerical integration of the above master equations.


\begin{thebibliography}{59}
\expandafter\ifx\csname natexlab\endcsname\relax\def\natexlab#1{#1}\fi
\expandafter\ifx\csname bibnamefont\endcsname\relax
  \def\bibnamefont#1{#1}\fi
\expandafter\ifx\csname bibfnamefont\endcsname\relax
  \def\bibfnamefont#1{#1}\fi
\expandafter\ifx\csname citenamefont\endcsname\relax
  \def\citenamefont#1{#1}\fi
\expandafter\ifx\csname url\endcsname\relax
  \def\url#1{\texttt{#1}}\fi
\expandafter\ifx\csname urlprefix\endcsname\relax\def\urlprefix{URL }\fi
\providecommand{\bibinfo}[2]{#2}
\providecommand{\eprint}[2][]{\url{#2}}

\bibitem[{\citenamefont{Castellano et~al.}(2009)\citenamefont{Castellano,
  Fortunato, and Loreto}}]{2009castellanoFL}
\bibinfo{author}{\bibfnamefont{C.}~\bibnamefont{Castellano}},
  \bibinfo{author}{\bibfnamefont{S.}~\bibnamefont{Fortunato}},
  \bibnamefont{and} \bibinfo{author}{\bibfnamefont{V.}~\bibnamefont{Loreto}},
  \bibinfo{journal}{Rev. Mod. Phys.} \textbf{\bibinfo{volume}{81}},
  \bibinfo{pages}{591} (\bibinfo{year}{2009}).

\bibitem[{\citenamefont{Galam}(2012)}]{2012galam}
\bibinfo{author}{\bibfnamefont{S.}~\bibnamefont{Galam}},
  \emph{\bibinfo{title}{Sociophysics: a physicist's modeling of
  psycho-political phenomena}} (\bibinfo{publisher}{Springer Science \&
  Business Media}, \bibinfo{year}{2012}).

\bibitem[{\citenamefont{Galam}(2008)}]{2008galam}
\bibinfo{author}{\bibfnamefont{S.}~\bibnamefont{Galam}},
  \bibinfo{journal}{International Journal of Modern Physics C}
  \textbf{\bibinfo{volume}{19}}, \bibinfo{pages}{409} (\bibinfo{year}{2008}).

\bibitem[{\citenamefont{Pan et~al.}(2017)\citenamefont{Pan, Qin, Xu, Tong, and
  He}}]{2017panQXTH}
\bibinfo{author}{\bibfnamefont{Q.}~\bibnamefont{Pan}},
  \bibinfo{author}{\bibfnamefont{Y.}~\bibnamefont{Qin}},
  \bibinfo{author}{\bibfnamefont{Y.}~\bibnamefont{Xu}},
  \bibinfo{author}{\bibfnamefont{M.}~\bibnamefont{Tong}}, \bibnamefont{and}
  \bibinfo{author}{\bibfnamefont{M.}~\bibnamefont{He}},
  \bibinfo{journal}{International Journal of Modern Physics C}
  \textbf{\bibinfo{volume}{28}}, \bibinfo{pages}{1750003}
  (\bibinfo{year}{2017}).

\bibitem[{\citenamefont{Javarone}(2014)}]{2014javarone}
\bibinfo{author}{\bibfnamefont{M.~A.} \bibnamefont{Javarone}},
  \bibinfo{journal}{Physica A: Statistical Mechanics and its Applications}
  \textbf{\bibinfo{volume}{414}}, \bibinfo{pages}{19} (\bibinfo{year}{2014}),
  ISSN \bibinfo{issn}{0378-4371}.

\bibitem[{\citenamefont{Javarone and Squartini}(2015)}]{2015javaroneS}
\bibinfo{author}{\bibfnamefont{M.~A.} \bibnamefont{Javarone}} \bibnamefont{and}
  \bibinfo{author}{\bibfnamefont{T.}~\bibnamefont{Squartini}},
  \bibinfo{journal}{Journal of Statistical Mechanics: Theory and Experiment}
  \textbf{\bibinfo{volume}{2015}}, \bibinfo{pages}{P10002}
  (\bibinfo{year}{2015}).

\bibitem[{\citenamefont{Crokidakis and Anteneodo}(2012)}]{2012crokidakisA}
\bibinfo{author}{\bibfnamefont{N.}~\bibnamefont{Crokidakis}} \bibnamefont{and}
  \bibinfo{author}{\bibfnamefont{C.}~\bibnamefont{Anteneodo}},
  \bibinfo{journal}{Phys. Rev. E} \textbf{\bibinfo{volume}{86}},
  \bibinfo{pages}{061127} (\bibinfo{year}{2012}).

\bibitem[{\citenamefont{Terranova et~al.}(2014)\citenamefont{Terranova,
  Revelli, and Sibona}}]{2014terranovaRS}
\bibinfo{author}{\bibfnamefont{G.~R.} \bibnamefont{Terranova}},
  \bibinfo{author}{\bibfnamefont{J.~A.} \bibnamefont{Revelli}},
  \bibnamefont{and} \bibinfo{author}{\bibfnamefont{G.~J.}
  \bibnamefont{Sibona}}, \bibinfo{journal}{EPL (Europhysics Letters)}
  \textbf{\bibinfo{volume}{105}}, \bibinfo{pages}{30007}
  (\bibinfo{year}{2014}).

\bibitem[{\citenamefont{Biswas}(2011)}]{2011biswas}
\bibinfo{author}{\bibfnamefont{S.}~\bibnamefont{Biswas}},
  \bibinfo{journal}{Phys. Rev. E} \textbf{\bibinfo{volume}{84}},
  \bibinfo{pages}{056106} (\bibinfo{year}{2011}).

\bibitem[{\citenamefont{{Calv\~ao} et~al.}(2016)\citenamefont{{Calv\~ao},
  Ramos, and Anteneodo}}]{2016calvaoRA}
\bibinfo{author}{\bibfnamefont{A.~M.} \bibnamefont{{Calv\~ao}}},
  \bibinfo{author}{\bibfnamefont{M.}~\bibnamefont{Ramos}}, \bibnamefont{and}
  \bibinfo{author}{\bibfnamefont{C.}~\bibnamefont{Anteneodo}},
  \bibinfo{journal}{Journal of Statistical Mechanics: Theory and Experiment}
  \textbf{\bibinfo{volume}{2016}}, \bibinfo{pages}{023405}
  (\bibinfo{year}{2016}).

\bibitem[{\citenamefont{Mukherjee and Chatterjee}(2016)}]{2016mukherjeeC}
\bibinfo{author}{\bibfnamefont{S.}~\bibnamefont{Mukherjee}} \bibnamefont{and}
  \bibinfo{author}{\bibfnamefont{A.}~\bibnamefont{Chatterjee}},
  \bibinfo{journal}{Phys. Rev. E} \textbf{\bibinfo{volume}{94}},
  \bibinfo{pages}{062317} (\bibinfo{year}{2016}).

\bibitem[{\citenamefont{Ramos et~al.}(2015)\citenamefont{Ramos, Shao, Reis,
  Anteneodo, Andrade, Havlin, and Makse}}]{2015ramosSRA}
\bibinfo{author}{\bibfnamefont{M.}~\bibnamefont{Ramos}},
  \bibinfo{author}{\bibfnamefont{J.}~\bibnamefont{Shao}},
  \bibinfo{author}{\bibfnamefont{S.~D.~S.} \bibnamefont{Reis}},
  \bibinfo{author}{\bibfnamefont{C.}~\bibnamefont{Anteneodo}},
  \bibinfo{author}{\bibfnamefont{J.~S.} \bibnamefont{Andrade}},
  \bibinfo{author}{\bibfnamefont{S.}~\bibnamefont{Havlin}}, \bibnamefont{and}
  \bibinfo{author}{\bibfnamefont{H.~A.} \bibnamefont{Makse}},
  \bibinfo{journal}{Scientific Reports} \textbf{\bibinfo{volume}{5}},
  \bibinfo{pages}{10032} (\bibinfo{year}{2015}).

\bibitem[{\citenamefont{Vieira and Crokidakis}(2016)}]{2016vieiraC}
\bibinfo{author}{\bibfnamefont{A.~R.} \bibnamefont{Vieira}} \bibnamefont{and}
  \bibinfo{author}{\bibfnamefont{N.}~\bibnamefont{Crokidakis}},
  \bibinfo{journal}{Physica A: Statistical Mechanics and its Applications}
  \textbf{\bibinfo{volume}{450}}, \bibinfo{pages}{30} (\bibinfo{year}{2016}).

\bibitem[{\citenamefont{Deffuant et~al.}(2000)\citenamefont{Deffuant, Neau,
  Amblard, and Weisbuch}}]{2000deffuantNAW}
\bibinfo{author}{\bibfnamefont{G.}~\bibnamefont{Deffuant}},
  \bibinfo{author}{\bibfnamefont{D.}~\bibnamefont{Neau}},
  \bibinfo{author}{\bibfnamefont{F.}~\bibnamefont{Amblard}}, \bibnamefont{and}
  \bibinfo{author}{\bibfnamefont{G.}~\bibnamefont{Weisbuch}},
  \bibinfo{journal}{Advances in Complex Systems} \textbf{\bibinfo{volume}{03}},
  \bibinfo{pages}{87} (\bibinfo{year}{2000}).

\bibitem[{\citenamefont{Lorenz}(2007)}]{2007lorenz}
\bibinfo{author}{\bibfnamefont{J.}~\bibnamefont{Lorenz}},
  \bibinfo{journal}{International Journal of Modern Physics C}
  \textbf{\bibinfo{volume}{18}}, \bibinfo{pages}{1819} (\bibinfo{year}{2007}).

\bibitem[{\citenamefont{Martins}(2008)}]{2008martins}
\bibinfo{author}{\bibfnamefont{A.~C.~R.} \bibnamefont{Martins}},
  \bibinfo{journal}{International Journal of Modern Physics C}
  \textbf{\bibinfo{volume}{19}}, \bibinfo{pages}{617} (\bibinfo{year}{2008}).

\bibitem[{\citenamefont{Lallouache et~al.}(2010)\citenamefont{Lallouache,
  Chakrabarti, Chakraborti, and Chakrabarti}}]{2010lallouacheCCC}
\bibinfo{author}{\bibfnamefont{M.}~\bibnamefont{Lallouache}},
  \bibinfo{author}{\bibfnamefont{A.~S.} \bibnamefont{Chakrabarti}},
  \bibinfo{author}{\bibfnamefont{A.}~\bibnamefont{Chakraborti}},
  \bibnamefont{and} \bibinfo{author}{\bibfnamefont{B.~K.}
  \bibnamefont{Chakrabarti}}, \bibinfo{journal}{Phys. Rev. E}
  \textbf{\bibinfo{volume}{82}}, \bibinfo{pages}{056112}
  (\bibinfo{year}{2010}).

\bibitem[{\citenamefont{Biswas et~al.}(2011)\citenamefont{Biswas, Chandra,
  Chatterjee, and Chakrabarti}}]{2011biswasCC}
\bibinfo{author}{\bibfnamefont{S.}~\bibnamefont{Biswas}},
  \bibinfo{author}{\bibfnamefont{A.~K.} \bibnamefont{Chandra}},
  \bibinfo{author}{\bibfnamefont{A.}~\bibnamefont{Chatterjee}},
  \bibnamefont{and} \bibinfo{author}{\bibfnamefont{B.~K.}
  \bibnamefont{Chakrabarti}}, in \emph{\bibinfo{booktitle}{Journal of physics:
  conference series}} (\bibinfo{organization}{IOP Publishing},
  \bibinfo{year}{2011}), vol. \bibinfo{volume}{297}, p.
  \bibinfo{pages}{012004}.

\bibitem[{\citenamefont{Vieira et~al.}(2016)\citenamefont{Vieira, Anteneodo,
  and Crokidakis}}]{2016vieiraAC}
\bibinfo{author}{\bibfnamefont{A.~R.} \bibnamefont{Vieira}},
  \bibinfo{author}{\bibfnamefont{C.}~\bibnamefont{Anteneodo}},
  \bibnamefont{and}
  \bibinfo{author}{\bibfnamefont{N.}~\bibnamefont{Crokidakis}},
  \bibinfo{journal}{Journal of Statistical Mechanics: Theory and Experiment}
  \textbf{\bibinfo{volume}{2016}}, \bibinfo{pages}{023204}
  (\bibinfo{year}{2016}).

\bibitem[{\citenamefont{Anteneodo and Crokidakis}(2017)}]{2017anteneodoC}
\bibinfo{author}{\bibfnamefont{C.}~\bibnamefont{Anteneodo}} \bibnamefont{and}
  \bibinfo{author}{\bibfnamefont{N.}~\bibnamefont{Crokidakis}},
  \bibinfo{journal}{Phys. Rev. E} \textbf{\bibinfo{volume}{95}},
  \bibinfo{pages}{042308} (\bibinfo{year}{2017}).

\bibitem[{\citenamefont{Biswas et~al.}(2012)\citenamefont{Biswas, Chatterjee,
  and Sen}}]{2012biswasCS}
\bibinfo{author}{\bibfnamefont{S.}~\bibnamefont{Biswas}},
  \bibinfo{author}{\bibfnamefont{A.}~\bibnamefont{Chatterjee}},
  \bibnamefont{and} \bibinfo{author}{\bibfnamefont{P.}~\bibnamefont{Sen}},
  \bibinfo{journal}{Physica A: Statistical Mechanics and its Applications}
  \textbf{\bibinfo{volume}{391}}, \bibinfo{pages}{3257 }
  (\bibinfo{year}{2012}), ISSN \bibinfo{issn}{0378-4371}.

\bibitem[{\citenamefont{Newman}(2006)}]{2006newman}
\bibinfo{author}{\bibfnamefont{M.~E.~J.} \bibnamefont{Newman}},
  \bibinfo{journal}{Proceedings of the National Academy of Sciences}
  \textbf{\bibinfo{volume}{103}}, \bibinfo{pages}{8577} (\bibinfo{year}{2006}),
  ISSN \bibinfo{issn}{0027-8424}.

\bibitem[{\citenamefont{Fortunato}(2010)}]{2010fortunato}
\bibinfo{author}{\bibfnamefont{S.}~\bibnamefont{Fortunato}},
  \bibinfo{journal}{Physics Reports} \textbf{\bibinfo{volume}{486}},
  \bibinfo{pages}{75 } (\bibinfo{year}{2010}), ISSN \bibinfo{issn}{0370-1573}.

\bibitem[{\citenamefont{Arenas et~al.}(2006)\citenamefont{Arenas,
  D\'{\i}az-Guilera, and P\'erez-Vicente}}]{2006arenasDPC}
\bibinfo{author}{\bibfnamefont{A.}~\bibnamefont{Arenas}},
  \bibinfo{author}{\bibfnamefont{A.}~\bibnamefont{D\'{\i}az-Guilera}},
  \bibnamefont{and} \bibinfo{author}{\bibfnamefont{C.~J.}
  \bibnamefont{P\'erez-Vicente}}, \bibinfo{journal}{Phys. Rev. Lett.}
  \textbf{\bibinfo{volume}{96}}, \bibinfo{pages}{114102}
  (\bibinfo{year}{2006}).

\bibitem[{\citenamefont{Li et~al.}(2008)\citenamefont{Li, Leyva, Almendral,
  Sendi\~na Nadal, Buld\'u, Havlin, and Boccaletti}}]{2008liLAS}
\bibinfo{author}{\bibfnamefont{D.}~\bibnamefont{Li}},
  \bibinfo{author}{\bibfnamefont{I.}~\bibnamefont{Leyva}},
  \bibinfo{author}{\bibfnamefont{J.~A.} \bibnamefont{Almendral}},
  \bibinfo{author}{\bibfnamefont{I.}~\bibnamefont{Sendi\~na Nadal}},
  \bibinfo{author}{\bibfnamefont{J.~M.} \bibnamefont{Buld\'u}},
  \bibinfo{author}{\bibfnamefont{S.}~\bibnamefont{Havlin}}, \bibnamefont{and}
  \bibinfo{author}{\bibfnamefont{S.}~\bibnamefont{Boccaletti}},
  \bibinfo{journal}{Phys. Rev. Lett.} \textbf{\bibinfo{volume}{101}},
  \bibinfo{pages}{168701} (\bibinfo{year}{2008}).

\bibitem[{\citenamefont{Liu and Hu}(2005)}]{2005liuH}
\bibinfo{author}{\bibfnamefont{Z.}~\bibnamefont{Liu}} \bibnamefont{and}
  \bibinfo{author}{\bibfnamefont{B.}~\bibnamefont{Hu}}, \bibinfo{journal}{EPL
  (Europhysics Letters)} \textbf{\bibinfo{volume}{72}}, \bibinfo{pages}{315}
  (\bibinfo{year}{2005}).

\bibitem[{\citenamefont{Huang et~al.}(2006)\citenamefont{Huang, Park, and
  Lai}}]{2006huangPL}
\bibinfo{author}{\bibfnamefont{L.}~\bibnamefont{Huang}},
  \bibinfo{author}{\bibfnamefont{K.}~\bibnamefont{Park}}, \bibnamefont{and}
  \bibinfo{author}{\bibfnamefont{Y.-C.} \bibnamefont{Lai}},
  \bibinfo{journal}{Phys. Rev. E} \textbf{\bibinfo{volume}{73}},
  \bibinfo{pages}{035103} (\bibinfo{year}{2006}).

\bibitem[{\citenamefont{Nematzadeh et~al.}(2014)\citenamefont{Nematzadeh,
  Ferrara, Flammini, and Ahn}}]{2014nematzadehFFA}
\bibinfo{author}{\bibfnamefont{A.}~\bibnamefont{Nematzadeh}},
  \bibinfo{author}{\bibfnamefont{E.}~\bibnamefont{Ferrara}},
  \bibinfo{author}{\bibfnamefont{A.}~\bibnamefont{Flammini}}, \bibnamefont{and}
  \bibinfo{author}{\bibfnamefont{Y.-Y.} \bibnamefont{Ahn}},
  \bibinfo{journal}{Phys. Rev. Lett.} \textbf{\bibinfo{volume}{113}},
  \bibinfo{pages}{088701} (\bibinfo{year}{2014}).

\bibitem[{\citenamefont{Lambiotte and Ausloos}(2007)}]{2007lambiotteA}
\bibinfo{author}{\bibfnamefont{R.}~\bibnamefont{Lambiotte}} \bibnamefont{and}
  \bibinfo{author}{\bibfnamefont{M.}~\bibnamefont{Ausloos}},
  \bibinfo{journal}{Journal of Statistical Mechanics: Theory and Experiment}
  \textbf{\bibinfo{volume}{2007}}, \bibinfo{pages}{P08026}
  (\bibinfo{year}{2007}).

\bibitem[{\citenamefont{Lambiotte et~al.}(2007)\citenamefont{Lambiotte,
  Ausloos, and Ho\l{}yst}}]{2007lambiotteAH}
\bibinfo{author}{\bibfnamefont{R.}~\bibnamefont{Lambiotte}},
  \bibinfo{author}{\bibfnamefont{M.}~\bibnamefont{Ausloos}}, \bibnamefont{and}
  \bibinfo{author}{\bibfnamefont{J.~A.} \bibnamefont{Ho\l{}yst}},
  \bibinfo{journal}{Phys. Rev. E} \textbf{\bibinfo{volume}{75}},
  \bibinfo{pages}{030101} (\bibinfo{year}{2007}).

\bibitem[{\citenamefont{Ru and Li-Ping}(2008)}]{2008ruL}
\bibinfo{author}{\bibfnamefont{W.}~\bibnamefont{Ru}} \bibnamefont{and}
  \bibinfo{author}{\bibfnamefont{C.}~\bibnamefont{Li-Ping}},
  \bibinfo{journal}{Chinese Physics Letters} \textbf{\bibinfo{volume}{25}},
  \bibinfo{pages}{1502} (\bibinfo{year}{2008}).

\bibitem[{\citenamefont{Si et~al.}(2009)\citenamefont{Si, Liu, and
  Zhang}}]{2009siLZ}
\bibinfo{author}{\bibfnamefont{X.}~\bibnamefont{Si}},
  \bibinfo{author}{\bibfnamefont{Y.}~\bibnamefont{Liu}}, \bibnamefont{and}
  \bibinfo{author}{\bibfnamefont{Z.}~\bibnamefont{Zhang}},
  \bibinfo{journal}{International Journal of Modern Physics C}
  \textbf{\bibinfo{volume}{20}}, \bibinfo{pages}{2013} (\bibinfo{year}{2009}).

\bibitem[{\citenamefont{Feng et~al.}(2015)\citenamefont{Feng, Han-Shuang, and
  Chuan-Sheng}}]{2015fengHC}
\bibinfo{author}{\bibfnamefont{H.}~\bibnamefont{Feng}},
  \bibinfo{author}{\bibfnamefont{C.}~\bibnamefont{Han-Shuang}},
  \bibnamefont{and}
  \bibinfo{author}{\bibfnamefont{S.}~\bibnamefont{Chuan-Sheng}},
  \bibinfo{journal}{Chinese Physics Letters} \textbf{\bibinfo{volume}{32}},
  \bibinfo{pages}{118902} (\bibinfo{year}{2015}).

\bibitem[{\citenamefont{Pan and Sinha}(2009)}]{2009panS}
\bibinfo{author}{\bibfnamefont{R.~K.} \bibnamefont{Pan}} \bibnamefont{and}
  \bibinfo{author}{\bibfnamefont{S.}~\bibnamefont{Sinha}},
  \bibinfo{journal}{EPL (Europhysics Letters)} \textbf{\bibinfo{volume}{85}},
  \bibinfo{pages}{68006} (\bibinfo{year}{2009}).

\bibitem[{\citenamefont{Dasgupta et~al.}(2009)\citenamefont{Dasgupta, Pan, and
  Sinha}}]{2009dasguptaPS}
\bibinfo{author}{\bibfnamefont{S.}~\bibnamefont{Dasgupta}},
  \bibinfo{author}{\bibfnamefont{R.~K.} \bibnamefont{Pan}}, \bibnamefont{and}
  \bibinfo{author}{\bibfnamefont{S.}~\bibnamefont{Sinha}},
  \bibinfo{journal}{Phys. Rev. E} \textbf{\bibinfo{volume}{80}},
  \bibinfo{pages}{025101} (\bibinfo{year}{2009}).

\bibitem[{\citenamefont{Suchecki and Ho\l{}yst}(2009)}]{2009sucheckiH}
\bibinfo{author}{\bibfnamefont{K.}~\bibnamefont{Suchecki}} \bibnamefont{and}
  \bibinfo{author}{\bibfnamefont{J.~A.} \bibnamefont{Ho\l{}yst}},
  \bibinfo{journal}{Phys. Rev. E} \textbf{\bibinfo{volume}{80}},
  \bibinfo{pages}{031110} (\bibinfo{year}{2009}).

\bibitem[{\citenamefont{Chen and Hou}(2011)}]{2011chenH}
\bibinfo{author}{\bibfnamefont{H.}~\bibnamefont{Chen}} \bibnamefont{and}
  \bibinfo{author}{\bibfnamefont{Z.}~\bibnamefont{Hou}},
  \bibinfo{journal}{Phys. Rev. E} \textbf{\bibinfo{volume}{83}},
  \bibinfo{pages}{046124} (\bibinfo{year}{2011}).

\bibitem[{\citenamefont{Conover et~al.}(2011)\citenamefont{Conover, Ratkiewicz,
  Francisco, Gon{\c{c}}alves, Menczer, and Flammini}}]{2011conoverRFG}
\bibinfo{author}{\bibfnamefont{M.~D.} \bibnamefont{Conover}},
  \bibinfo{author}{\bibfnamefont{J.}~\bibnamefont{Ratkiewicz}},
  \bibinfo{author}{\bibfnamefont{M.}~\bibnamefont{Francisco}},
  \bibinfo{author}{\bibfnamefont{B.}~\bibnamefont{Gon{\c{c}}alves}},
  \bibinfo{author}{\bibfnamefont{F.}~\bibnamefont{Menczer}}, \bibnamefont{and}
  \bibinfo{author}{\bibfnamefont{A.}~\bibnamefont{Flammini}}, in
  \emph{\bibinfo{booktitle}{Fifth international AAAI conference on weblogs and
  social media}} (\bibinfo{year}{2011}).

\bibitem[{\citenamefont{{Cota} et~al.}(2019)\citenamefont{{Cota}, {Ferreira},
  {Pastor-Satorras}, and {Starnini}}}]{2019cotaFPS}
\bibinfo{author}{\bibfnamefont{W.}~\bibnamefont{{Cota}}},
  \bibinfo{author}{\bibfnamefont{S.~C.} \bibnamefont{{Ferreira}}},
  \bibinfo{author}{\bibfnamefont{R.}~\bibnamefont{{Pastor-Satorras}}},
  \bibnamefont{and}
  \bibinfo{author}{\bibfnamefont{M.}~\bibnamefont{{Starnini}}},
  \bibinfo{journal}{arXiv e-prints} \bibinfo{eid}{arXiv:1901.03688}
  (\bibinfo{year}{2019}), \eprint{1901.03688}.

\bibitem[{\citenamefont{Yardi and Boyd}(2010)}]{2010yardiB}
\bibinfo{author}{\bibfnamefont{S.}~\bibnamefont{Yardi}} \bibnamefont{and}
  \bibinfo{author}{\bibfnamefont{D.}~\bibnamefont{Boyd}},
  \bibinfo{journal}{Bulletin of Science, Technology {\&} Society}
  \textbf{\bibinfo{volume}{30}}, \bibinfo{pages}{316} (\bibinfo{year}{2010}).

\bibitem[{\citenamefont{Hewstone et~al.}(2002)\citenamefont{Hewstone, Rubin,
  and Willis}}]{2002hewstoneRW}
\bibinfo{author}{\bibfnamefont{M.}~\bibnamefont{Hewstone}},
  \bibinfo{author}{\bibfnamefont{M.}~\bibnamefont{Rubin}}, \bibnamefont{and}
  \bibinfo{author}{\bibfnamefont{H.}~\bibnamefont{Willis}},
  \bibinfo{journal}{Annual Review of Psychology} \textbf{\bibinfo{volume}{53}},
  \bibinfo{pages}{575} (\bibinfo{year}{2002}), \bibinfo{note}{pMID: 11752497}.

\bibitem[{\citenamefont{Efferson et~al.}(2008)\citenamefont{Efferson, Lalive,
  and Fehr}}]{2008effersonLF}
\bibinfo{author}{\bibfnamefont{C.}~\bibnamefont{Efferson}},
  \bibinfo{author}{\bibfnamefont{R.}~\bibnamefont{Lalive}}, \bibnamefont{and}
  \bibinfo{author}{\bibfnamefont{E.}~\bibnamefont{Fehr}},
  \bibinfo{journal}{Science} \textbf{\bibinfo{volume}{321}},
  \bibinfo{pages}{1844} (\bibinfo{year}{2008}), ISSN \bibinfo{issn}{0036-8075}.

\bibitem[{\citenamefont{Galam}(2002)}]{2002galam}
\bibinfo{author}{\bibfnamefont{S.}~\bibnamefont{Galam}}, \bibinfo{journal}{The
  European Physical Journal B} \textbf{\bibinfo{volume}{25}},
  \bibinfo{pages}{403} (\bibinfo{year}{2002}).

\bibitem[{\citenamefont{Biswas and Sen}(2017)}]{2017biswasS}
\bibinfo{author}{\bibfnamefont{S.}~\bibnamefont{Biswas}} \bibnamefont{and}
  \bibinfo{author}{\bibfnamefont{P.}~\bibnamefont{Sen}},
  \bibinfo{journal}{Physical Review E} \textbf{\bibinfo{volume}{96}},
  \bibinfo{pages}{032303} (\bibinfo{year}{2017}).

\bibitem[{\citenamefont{Galam and Jacobs}(2007)}]{2007galamJ}
\bibinfo{author}{\bibfnamefont{S.}~\bibnamefont{Galam}} \bibnamefont{and}
  \bibinfo{author}{\bibfnamefont{F.}~\bibnamefont{Jacobs}},
  \bibinfo{journal}{Physica A: Statistical Mechanics and its Applications}
  \textbf{\bibinfo{volume}{381}}, \bibinfo{pages}{366} (\bibinfo{year}{2007}).

\bibitem[{\citenamefont{Girvan and Newman}(2002)}]{2002girvanN}
\bibinfo{author}{\bibfnamefont{M.}~\bibnamefont{Girvan}} \bibnamefont{and}
  \bibinfo{author}{\bibfnamefont{M.~E.~J.} \bibnamefont{Newman}},
  \bibinfo{journal}{Proceedings of the National Academy of Sciences}
  \textbf{\bibinfo{volume}{99}}, \bibinfo{pages}{7821} (\bibinfo{year}{2002}),
  ISSN \bibinfo{issn}{0027-8424}.

\bibitem[{\citenamefont{Lancichinetti et~al.}(2008)\citenamefont{Lancichinetti,
  Fortunato, and Radicchi}}]{2008lancichinettiFR}
\bibinfo{author}{\bibfnamefont{A.}~\bibnamefont{Lancichinetti}},
  \bibinfo{author}{\bibfnamefont{S.}~\bibnamefont{Fortunato}},
  \bibnamefont{and} \bibinfo{author}{\bibfnamefont{F.}~\bibnamefont{Radicchi}},
  \bibinfo{journal}{Phys. Rev. E} \textbf{\bibinfo{volume}{78}},
  \bibinfo{pages}{046110} (\bibinfo{year}{2008}).

\bibitem[{\citenamefont{Karrer and Newman}(2011)}]{2011karrerN}
\bibinfo{author}{\bibfnamefont{B.}~\bibnamefont{Karrer}} \bibnamefont{and}
  \bibinfo{author}{\bibfnamefont{M.~E.~J.} \bibnamefont{Newman}},
  \bibinfo{journal}{Phys. Rev. E} \textbf{\bibinfo{volume}{83}},
  \bibinfo{pages}{016107} (\bibinfo{year}{2011}).

\bibitem[{\citenamefont{Javarone and Marinazzo}(2018)}]{2018javaroneM}
\bibinfo{author}{\bibfnamefont{M.~A.} \bibnamefont{Javarone}} \bibnamefont{and}
  \bibinfo{author}{\bibfnamefont{D.}~\bibnamefont{Marinazzo}},
  \bibinfo{journal}{Complexity} \textbf{\bibinfo{volume}{2018}},
  \bibinfo{pages}{1} (\bibinfo{year}{2018}).

\bibitem[{\citenamefont{Encinas et~al.}(2018)\citenamefont{Encinas, Harunari,
  de~Oliveira, and Fiore}}]{2018encinasHOF}
\bibinfo{author}{\bibfnamefont{J.~M.} \bibnamefont{Encinas}},
  \bibinfo{author}{\bibfnamefont{P.~E.} \bibnamefont{Harunari}},
  \bibinfo{author}{\bibfnamefont{M.~M.} \bibnamefont{de~Oliveira}},
  \bibnamefont{and} \bibinfo{author}{\bibfnamefont{C.~E.} \bibnamefont{Fiore}},
  \bibinfo{journal}{Scientific Reports} \textbf{\bibinfo{volume}{8}},
  \bibinfo{pages}{9338} (\bibinfo{year}{2018}).

\bibitem[{\citenamefont{Jedrzejewski and
  Sznajd-Weron}(2018)}]{2018jedrzejewskiS}
\bibinfo{author}{\bibfnamefont{A.}~\bibnamefont{Jedrzejewski}}
  \bibnamefont{and}
  \bibinfo{author}{\bibfnamefont{K.}~\bibnamefont{Sznajd-Weron}},
  \bibinfo{journal}{Physica A: Statistical Mechanics and its Applications}
  \textbf{\bibinfo{volume}{505}}, \bibinfo{pages}{306 } (\bibinfo{year}{2018}),
  ISSN \bibinfo{issn}{0378-4371}.

\bibitem[{\citenamefont{Crokidakis}(2013)}]{2013crokidakis}
\bibinfo{author}{\bibfnamefont{N.}~\bibnamefont{Crokidakis}},
  \bibinfo{journal}{Journal of Statistical Mechanics: Theory and Experiment}
  \textbf{\bibinfo{volume}{2013}}, \bibinfo{pages}{P07008}
  (\bibinfo{year}{2013}).

\bibitem[{\citenamefont{Ben-Naim}(2005)}]{2005bennaim}
\bibinfo{author}{\bibfnamefont{E.}~\bibnamefont{Ben-Naim}},
  \bibinfo{journal}{Europhysics Letters ({EPL})} \textbf{\bibinfo{volume}{69}},
  \bibinfo{pages}{671} (\bibinfo{year}{2005}).

\bibitem[{\citenamefont{Huang et~al.}(2008)\citenamefont{Huang, Cao, Wang, and
  Qu}}]{2008huangCWQ}
\bibinfo{author}{\bibfnamefont{G.}~\bibnamefont{Huang}},
  \bibinfo{author}{\bibfnamefont{J.}~\bibnamefont{Cao}},
  \bibinfo{author}{\bibfnamefont{G.}~\bibnamefont{Wang}}, \bibnamefont{and}
  \bibinfo{author}{\bibfnamefont{Y.}~\bibnamefont{Qu}},
  \bibinfo{journal}{Physica A: Statistical Mechanics and its Applications}
  \textbf{\bibinfo{volume}{387}}, \bibinfo{pages}{4665} (\bibinfo{year}{2008}).

\bibitem[{\citenamefont{Huang et~al.}(2009)\citenamefont{Huang, Cao, and
  Qu}}]{2009huangCQ}
\bibinfo{author}{\bibfnamefont{G.}~\bibnamefont{Huang}},
  \bibinfo{author}{\bibfnamefont{J.}~\bibnamefont{Cao}}, \bibnamefont{and}
  \bibinfo{author}{\bibfnamefont{Y.}~\bibnamefont{Qu}},
  \bibinfo{journal}{Physica A: Statistical Mechanics and its Applications}
  \textbf{\bibinfo{volume}{388}}, \bibinfo{pages}{3911} (\bibinfo{year}{2009}).

\bibitem[{\citenamefont{Shen and Liu}(2010)}]{2010shenL}
\bibinfo{author}{\bibfnamefont{B.}~\bibnamefont{Shen}} \bibnamefont{and}
  \bibinfo{author}{\bibfnamefont{Y.}~\bibnamefont{Liu}},
  \bibinfo{journal}{International Journal of Modern Physics C}
  \textbf{\bibinfo{volume}{21}}, \bibinfo{pages}{1001} (\bibinfo{year}{2010}).

\bibitem[{\citenamefont{Wu et~al.}(2017)\citenamefont{Wu, Xiong, and
  Zhang}}]{2017wuXZ}
\bibinfo{author}{\bibfnamefont{Y.}~\bibnamefont{Wu}},
  \bibinfo{author}{\bibfnamefont{X.}~\bibnamefont{Xiong}}, \bibnamefont{and}
  \bibinfo{author}{\bibfnamefont{Y.}~\bibnamefont{Zhang}},
  \bibinfo{journal}{Modern Physics Letters B} \textbf{\bibinfo{volume}{31}},
  \bibinfo{pages}{1750058} (\bibinfo{year}{2017}).

\bibitem[{\citenamefont{Crokidakis and de~Oliveira}(2014)}]{2014crokidakisO}
\bibinfo{author}{\bibfnamefont{N.}~\bibnamefont{Crokidakis}} \bibnamefont{and}
  \bibinfo{author}{\bibfnamefont{P.~M.~C.} \bibnamefont{de~Oliveira}},
  \bibinfo{journal}{Physica A: Statistical Mechanics and its Applications}
  \textbf{\bibinfo{volume}{409}}, \bibinfo{pages}{48} (\bibinfo{year}{2014}).

\bibitem[{\citenamefont{Pires et~al.}(2018)\citenamefont{Pires, Oestereich, and
  Crokidakis}}]{2018piresOC}
\bibinfo{author}{\bibfnamefont{M.~A.} \bibnamefont{Pires}},
  \bibinfo{author}{\bibfnamefont{A.~L.} \bibnamefont{Oestereich}},
  \bibnamefont{and}
  \bibinfo{author}{\bibfnamefont{N.}~\bibnamefont{Crokidakis}},
  \bibinfo{journal}{Journal of Statistical Mechanics: Theory and Experiment}
  \textbf{\bibinfo{volume}{2018}}, \bibinfo{pages}{053407}
  (\bibinfo{year}{2018}).

\end{thebibliography}

\end{document}